\documentclass[prb,aps,twocolumn,groupedaddress,float,showpacs,final,superscriptaddress]{revtex4-1}

\usepackage[latin3]{inputenc}
\usepackage[makeroom]{cancel}
\usepackage{subfigure}
\usepackage{graphicx}
\usepackage{amsmath}
\usepackage{amsfonts}
\usepackage{amssymb}
\usepackage{color}
\usepackage[left=2cm,right=2cm,top=2cm,bottom=2cm]{geometry}
\usepackage[export]{adjustbox}
\usepackage{graphicx}
\usepackage{dcolumn}
\usepackage{bm}
\usepackage{simplewick}
\usepackage{array}
\usepackage{appendix}
\usepackage{relsize}
\usepackage{footnote}

\newcommand{\be}{\begin{equation}}
	\newcommand{\ee}{\end{equation}}
\newcommand{\beq}{\begin{equation}}
	\newcommand{\eeq}{\end{equation}}
\newcommand{\bea}{\begin{eqnarray}}
	\newcommand{\eea}{\end{eqnarray}}

\RequirePackage[
   hyperindex,colorlinks,bookmarksnumbered,
   plainpages=true,pdfstartview=FitH]{hyperref}
\hypersetup{linkcolor=blue,urlcolor=blue,citecolor=blue}

\renewcommand{\phi}{\varphi}
\renewcommand{\epsilon}{\varepsilon}

\renewcommand{\vec}[1]{{\bf #1}}

\providecommand{\U}[1]{\protect\rule{.1in}{.1in}}
\providecommand{\U}[1]{\protect\rule{.1in}{.1in}}

\begin{document}
\title{Collision-dominated conductance in clean 2D metals}
\author{A. Uzair}
\affiliation{The Abdus Salam International Center for Theoretical Physics, Strada Costiera 11, 34151 Trieste, Italy.}
\affiliation{Department of Physics, Quaid-i-Azam University, Islamabad 45320, Pakistan.}
\affiliation{National Centre for Physics, Islamabad 44000, Pakistan.}
\author{K. Sabeeh}
\affiliation{Department of Physics, Quaid-i-Azam University, Islamabad 45320, Pakistan.}
\author{Markus M\"uller}
\email{{markus.mueller@psi.ch}}
\affiliation{The Abdus Salam International Center for Theoretical Physics, Strada Costiera 11, 34151 Trieste, Italy.}
\affiliation{Paul Scherrer Institute, Condensed Matter Theory, PSI Villigen, Switzerland.}
\affiliation{Department of Physics, University of Basel, Klingelbergstrasse 82, CH-4056 Basel, Switzerland.}
\begin{abstract}
We study the temperature-dependent corrections to the conductance due to electron-electron ({\em e-e}) interactions in clean two-dimensional conductors, such as lightly doped graphene or other Dirac matter. 
We use  semiclassical Boltzmann kinetic theory to solve the problem of collision-dominated transport between reflection-free contacts.
Time-reversal symmetry and the kinematic constraints of scattering in two dimensions (2D) ensure that inversion-odd and inversion-even distortions of the quasiparticle distribution relax with parametrically different rates at low temperature. This entails the surprising result that at lowest temperatures the conductance of very long samples tends to the noninteracting, ballistic conductance, despite the relaxation of the quasiparticle distribution to a drifting equilibrium. The relative correction to the conductance  depends on the ratio of relaxation rates of even and odd modes and scales as $\delta G/G_{\rm ballistic}\sim\left(T/\varepsilon_F\right) \sqrt{\log\left(\frac{\varepsilon_F}{T}\right)}$, in stark contrast to the behavior in other dimensionalities. This holds generally in 2D systems with simply connected and convex but otherwise arbitrary Fermi surfaces, as long as {\em e-e} scattering processes are dominant and umklapp scattering is negligible. These results are especially relevant to the bulk of wide and long suspended high-mobility graphene sheets.  
\end{abstract}
\startpage{01}
\endpage{02}
\author{}
\maketitle
\section*{Introduction}
Recent experiments on suspended single layer-graphene~\cite{b9,b16,b37} and on conventional two-dimensional electron gases (2DEG) \cite{b39,b40,b41,b47,b48,b49} have reached very high levels of purity and thus high mobilities. %\MM{check: Philipp Moll work?, Morpurgo latest suspended multilayer work? E. Andrei?} 
This has raised interesting questions as to the role of electron-electron interactions in such systems, especially in Dirac or Weyl materials such as graphene, as well as in other ultraclean 2D (semi)metals. Those systems are of particular interest because the electron-electron ({\em e-e}) interactions compete differently with the kinetic energy than in conventional materials with quadratic dispersion, where the interactions are significant only at low carrier density. In contrast, in graphene, on the 2D surfaces of three dimensional (3D) topological insulators, or on 2D system where the dispersion close to a Dirac or Weyl point at the Fermi level is linear, the interactions remain significant at all densities \cite{b6,b28}. Although the carrier concentration in a conventional 2DEG in semiconductor heterostructures can in principle be tuned down by gate voltage, it is  very challenging to exhibit the effect of {\em e-e} interactions on charge transport, because at the relevant low densities transport tends to be dominated by impurity scattering, e.g., from random charges at the interface between the 2DEG and the substrate \cite{b30,b31,b32}. The availability of very clean suspended 2D materials such as graphene, with mean free paths of the order of microns \cite{b9,b16,b37} and with significant {\em e-e} interactions, has thus made it possible to investigate regimes in which {\em e-e} scattering dominates all other scattering channels. This so-called collision-dominated regime occurs at intermediate temperatures. On one hand, the temperature has to be high enough such that the inelastic scattering rate due to {\em e-e} interactions, $\tau_{ee}^{-1}(T)$, is much greater than the elastic scattering rate due to impurities \cite{b30,b31,b32,b53,b54}. On the other hand, the temperature has to be low enough so that electron-phonon scattering \cite{b33,b34,b35,b51,b52} is still subdominant. In cases (such as in lightly doped graphene) where the structure of the Fermi surface and the dominant interactions are such as to suppress umklapp scattering \cite{b38,b62}, the total momentum of the electron fluid is essentially conserved by the translation-invariant {\em e-e} interactions. The electron fluid in the collision-dominated regime then behaves like a hydrodynamic fluid \cite{b58,b59,b60,prbshaffique,Moll1061}.\par
\subsection{Interaction-dominated , collective transport in electron systems}
The hydrodynamic description yields insight into the behavior of electronic current flow in the bulk of relatively large spatial structures with nontrivial geometries and boundary conditions%{\color{red}{Morpurgo, Moll, IBM, others? Levitov}} 
\cite{b43}. Lateral boundaries affect the transport properties by breaking the momentum conservation and providing a source of friction. In highly viscous fluids (low Reynolds numbers), nonlinear phenomena like turbulence do not arise, and the electron flow will be steady. In this case, the combined effect of momentum-conserving {\em e-e} and nonconserving diffusive boundary scattering~\cite{b44} leads to interesting phenomena such as the Gurzhi effect, an electronic analog of Poiseuille flow which has a counter-intuitive impact on the longitudinal conductivity. In this regard, superballistic flow exceeding upper bounds established for the conductance of noninteracting systems has been reported and was attributed to the collective motion of interacting electrons which reduces momentum loss at the boundary \cite{b44,b61,min,prbgurzhi3,prbjong}. Negative nonlocal voltages appearing near current injection points due to spatially inhomogeneous current flow and vorticity has been detected in recent experiments on graphene~\cite{b45,b46,b56,b57}. However, these anomalous interaction-dominated effects are prominent only within the viscous boundary layer; outside that layer, the flow is almost potential. In this article, we do not address such viscous effects, as we consider a wide sample with negligible impact from lateral boundaries. Our results are therefore valid for the bulk of the sample far from lateral boundaries where only momentum-conserving {\em e-e} scattering leads to quasiparticle scattering and equilibration. Instead, we address a situation where the hydrodynamic description fails, namely in the vicinity of the contacts through which the current enters and exits the sample, and where the electron liquid is far from equilibrium. We address the deceptively simple question: What is the conductance through a wide and very long strip, in the regime where transport is collision-dominated and momentum is conserved? Hydrodynamic considerations imply that far from the contacts and the boundaries, the electron liquid should reach a drifting equilibrium state with a drift velocity (and thus a current) proportional to the applied bias. However, these considerations do not allow one to determine the drift velocity as a function of the bias, since this requires the matching of the bulk of the sample to the boundary conditions at the contacts, which are outside the domain of validity of hydrodynamics.\par
\subsection{Conductance and conductivity in Dirac liquids}
In Dirac liquids, where interactions are significant at all densities, the question of interaction-dominated current flow is interesting for a wide range of densities. In general, as long as one can neglect momentum-degrading scatterings, one expects a finite conductance independent of the length of the strip (corresponding to infinite bulk conductivity, as expected from the $1/\omega $ pole in the frequency-dependent conductivity $\sigma(\omega)$ obtained in hydrodynamics). However, when the chemical potential lies at the Dirac point, a special situation arises, since there the electric current induced by an electric field will carry no total momentum, because of particle-hole symmetry. Therefore, the electrical current can relax despite the conservation of momentum~\cite{b8,b10}. In that case, a finite, interaction-dominated conductivity arises (with a conductance that decreases with inverse sample length according to Ohm's law), which was evaluated in Refs.~[\onlinecite{b7,b8,b10}]. Here instead, we restrict our investigation to the effect of interactions on the conductance of 2D systems at finite carrier density, away from the Dirac point. Our study is in part motivated by experiments on suspended graphene, where a length-independent conductance was reported in clean samples that were longer than the estimated inelastic scattering length \cite{b9,b16,b37}. However, our results apply equally well to the conductance in conventional systems with parabolic dispersion, as long as interactions provide the dominant scattering channel \cite{b21,b34}.\par
In the absence of interactions, in clean samples, electrons propagate ballistically through the system, keeping a memory of the lead they originate from. Essentially the same holds for weakly interacting systems which are much shorter than the inelastic scattering length, $l_{\mathrm{inel}}=v_{F}\tau_{ee}\left( T\right) $. In such short samples, in the presence of reflection-free contacts, the conductance is given by the standard Landauer-B\"uttiker formula, which sums the transmission probabilities of conducting modes \cite{b22,b23}. However, in longer samples of length $L>l_{\mathrm{inel}}$, the interactions modify the distribution function within the sample, resulting in additional resistance, and thus a decreasing conductance with increasing length~\cite{b2}. In general, the distribution function depends on the distance from the leads.  For long samples, $L\gg l_{\mathrm{inel}}$, and with interactions that conserve momentum, the distribution in the bulk will relax to a drifting equilibrium which is stable under collisions. The negative interaction correction to the ballistic conductance then saturates to a length-independent value. The drifting equilibrium is characterized by a nonzero drift (or center-of-mass) velocity $v_{d}$, which, within linear response, is proportional to the applied bias.\par
\subsection{Collision-dominated conductance: 1D vs. 2D}
The conductance of interaction-dominated electronic systems and the emergence of a drifting equilibrium state in long samples has previously been addressed for 1D systems \cite{b1,b2,b25,b26,b27}. Reference~[\onlinecite{b2}] studied the effect of interactions on the conductance in long wires, as well as the crossover between short ballistic samples and long interaction-dominated wires. An electron fluid in 1D is highly constrained by conservation laws, and it was found that the temperature dependence of the interaction-dominated transport could be determined without specifying the interactions in any detail but purely by exploiting those conservation laws. It was shown that in 1D, relaxation occurs by decreasing the imbalance between right and left movers, which involves multiple three-particle scattering events. In these systems with parabolic dispersion, the correction to the conductance was derived to be proportional to $\frac{L}{L+l_{\mathrm{eq}}}\left(\frac{T}{\mu}\right)^{2}$, where $l_{\mathrm{eq}}$ is an inelastic scattering or equilibration length which becomes exponentially large at low temperatures. The coefficient of the correction turned out to be the universal, interaction-independent number $\frac{\pi^2}{12}$.\par
Here, we address the analogous question for 2D samples which turns out to constitute a special and conceptually interesting case. In contrast to 1D, the scattering processes are far less restricted, and accordingly the solution is more complex. {\em A priori,} one could  expect the conductance to reflect the strength and the characteristics of the specific interactions. However, surprisingly, we find in this work that under certain general symmetry assumptions the interaction-induced decrease of the low-temperature conductance  invariably follows the temperature dependence $T/\varepsilon_F[\log(\varepsilon_F/T)]^{1/2}$, with a coefficient  {\em independent} of the strength of the interactions. The coefficient merely depends on aspects such as the shape of the Fermi surface and the range of the interactions.
% We identify the kinetically allowed two-particle scattering processes which are most efficient in equilibrating the electron fluid. At sufficiently low temperatures, for convex and simply connected but otherwise arbitrary 2D Fermi surfaces, there are two dominant relaxation processes \cite{b4}: head-on-collisions and forward scattering. Head-on-collisions, although logarithmically less frequent than forward scattering events in 2D, turn out to be the most efficient in relaxing the system towards an equilibrium distribution. This is because they allow arbitrarily large angle scattering. In contrast, forward scattering is restricted to small angle scattering, which induces relaxation of the energy distribution, restricted within given angular sectors, at logarithmically faster rates than global equilibration occurs \cite{b4,b19,b20,b50}.\par
This result strongly rests on the difference between even and odd (under momentum inversion) distortions of the  distribution of quasiparticles from equilibrium, which in $D=2$ dimensions have parametrically different relaxation rates at low temperature \cite{odd,prbgurzhi1}. Such a classification of modes is, however, meaningful only in situations where the microscopic scattering rates obey an inversion symmetry (invariance under $k_x\rightarrow-k_x$), as, for example, in systems with time-reversal symmetry and no significant spin-orbit coupling.
%, which is guaranteed by time reversal invariance. 
%{\color{red}{For a single Dirac cone in graphene, one further needs the  symmetry of the Fermi surface under $k\to 2K-k$, since time reversal  relates the Fermi surfaces around the $K$ and $K'$ points, while we are concerned with intra-cone scattering with small momentum transfer $\Delta k\ll K-K'$}} . 
Under these conditions, head-on-collisions \cite{b21} make even modes relax much faster than odd ones by an inverse power of temperature. This reflects in the temperature-dependent correction of the conductance, which scales as the square root of the ratio of the two relaxation rates.\par
These results apply to generic 2D systems with the aforementioned symmetries and convex Fermi surfaces. However, since {\em e-e} interaction effects are particularly pronounced in Dirac materials \cite{b6,b28}, these, and in particular suspended graphene of high mobility, are among the most promising candidates for experimentally detecting the predicted signatures of interactions in the conductance.\par
Our results are particular to 2D systems, because in 1D systems binary collisions do not relax the total momentum, while in higher dimensions, $D>2$, there exist additional scattering channels (other than head-on collisions), which relax even and odd modes essentially equally rapidly, independently of their parity under inversion of momenta. \par
The remainder of the paper is organized as follows. Section~\ref{M} introduces the model setup of a wide strip between biasing leads. In Sec.~\ref{BE}, we discuss the Boltzmann equation which governs transport when {\em e-e} interactions can be treated semi-classically. In Sec.~\ref{ISCO} we restrict  to the case where the collisions obey an inversion symmetry, and Sec.~\ref{Umklapp} shows how the collision-dominated ballistic conductance crosses over to Ohmic conduction in the presence of weak umklapp scattering. Section~\ref{RAV} discusses  how to reduce the problem  at low temperatures to purely angular degrees of freedom. Section~\ref{CSISS} is devoted to generic systems with inversion symmetric collision kernel. Section~\ref{RI} reduces to a rotationally invariant Fermi surface and provides explicit results that apply to graphene in particular. Section \ref{RD} discusses what crucial role the dimensionality $D=2$ plays for our results. We conclude with a summary and an outlook for future work. The appendix derives the parametrically differing relaxation rates for even and odd modes, respectively. 
\section{Collision-dominated conductance} \label{M}
Our aim is to calculate the corrections to the non-interacting, ballistic conductance that arise as a consequence of inelastic electron-electron scattering, but in the absence of impurity or phonon-scattering processes. The latter  do not conserve the momentum of the electron fluid and thus induce the decay of currents. This would establish a finite Ohmic conductivity and thus a  conductance vanishing as $1/L$ in the limit of long sample length $L$. If instead only translationally invariant interactions are present, the momentum of an electron fluid in homogeneous space would be conserved, so that the associated current cannot decay. However, the crystal lattice breaks the translational invariance and thus reduces the conservation of momentum to the conservation of quasimomentum, modulo reciprocal lattice vectors. However, the conservation of total quasimomentum $\sum_{\vec{k}\,\mathrm{occ}}\vec{k}$ of all quasiparticles will remain an excellent approximation if umklapp processes can be neglected at low temperatures. This is, for example, the case when the Fermi surface is small as compared to reciprocal lattice vectors, such that umklapp processes require high-energy excitations above $\varepsilon_F$ and thus are exponentially suppressed. Particularly interesting examples are  surfaces of topological insulators with a single Dirac point close  to the Fermi surface. Another example is lightly doped graphene, where umklapp processes only lead to a redistribution of particles in the vicinity of $K$ and $K'$ points, but cannot relax the total quasimomentum, whereby quasimomenta of quasiparticles are measured as the distance of the wave vector $\vec{k}$ from  the closer of the two Dirac points.\par
In what follows, we consider systems where umklapp processes are  negligible as compared to other scatterings induced by {\em e-e} interactions. Under such circumstances one expects a finite conductance to survive in the limit $L\to \infty$. 
\subsection{Model}
\begin{figure}[h]
	\begin{center}
		\includegraphics[width=\linewidth]{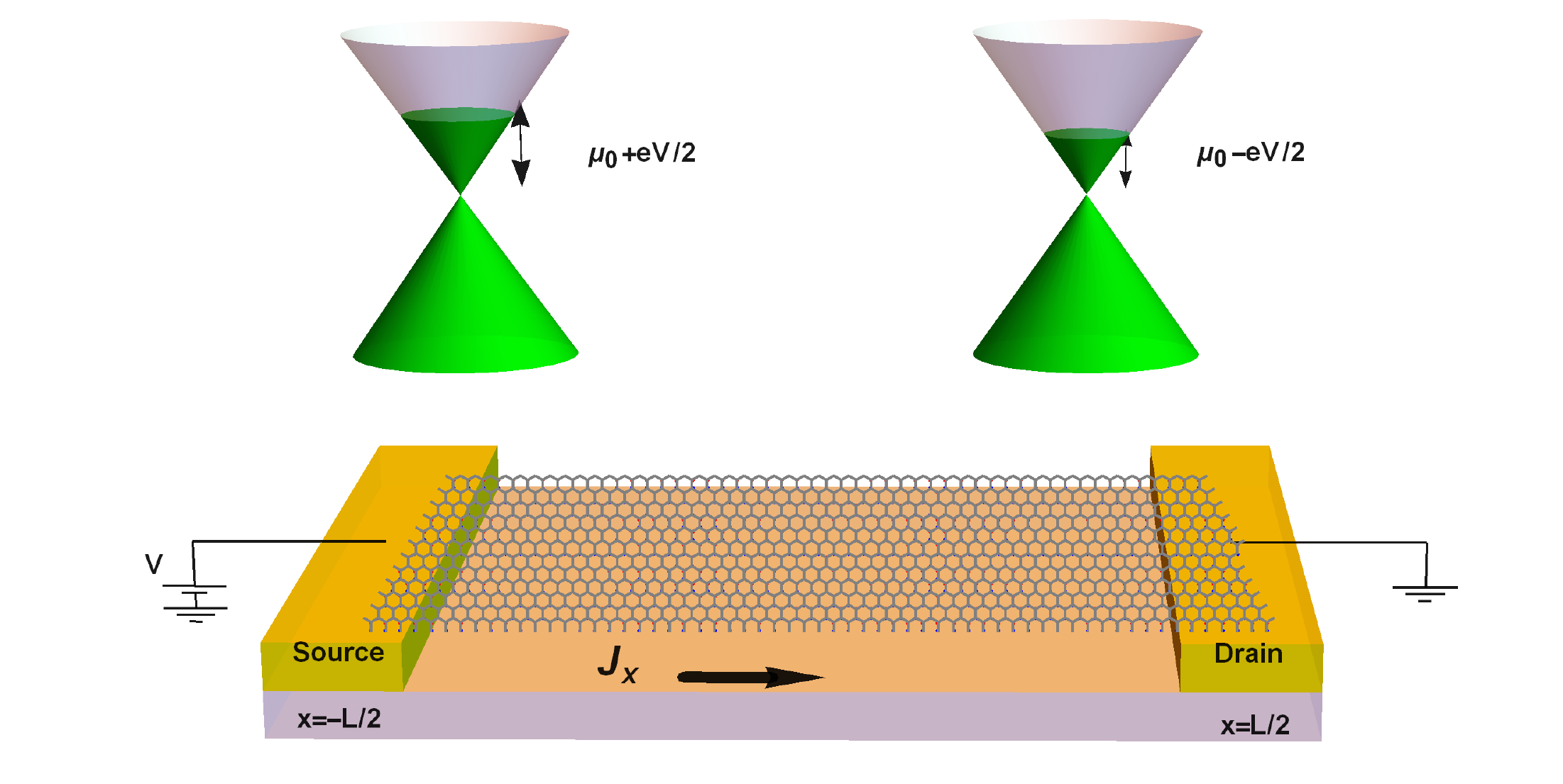} \vspace*{-0.1cm} 
	\end{center}
	\vspace*{-0.1cm}
	\caption{Free-hanging graphene sheet in the {\em x-y} plane. The leads are gated so as to have a finite Fermi surface, with quasi-particles  described by Fermi-liquid theory. A potential difference is applied along the $x$ axis via weakly coupled leads at $x=\pm\frac{L}{2}$. The bias results in a longitudinal current density $J_{x}$.}
	\label{fig:1}
\end{figure}

We consider an impurity-free, conducting 2D sample of infinite width $W\rightarrow\infty$ and length $L$, connected to two leads at $x=\pm\frac{L}{2}$ as illustrated in Fig. \ref{fig:1}. We assume the conductor to form a Fermi liquid. Such 2D liquids  arise, e.g., in the form of a conducting sheet of a metal, such as a suspended graphene sheet, or at a surface of a 3D topological insulator (with Dirac spectrum). Time-reversal symmetry ensures that $E_{\vec{k}}{=}E_{-\vec{k}}$ and thus $\vec{v(k)} = \partial E_{\vec{k}}/\partial \vec{k} {=}{-}\vec{v(-k)}$. To lowest order we will linearize the quasiparticle dispersion close to the Fermi energy, $E_{\vec{k}}{=}\hbar v_{F}(k-k_F)$ \cite{b6}, with Fermi velocity $v_{F}$. For simplicity, we assume a circularly symmetric Fermi surface and neglect the energy and angle dependence of the velocity.  To further simplify our analysis, we will not consider additional quantum numbers such as valley or spin index. It is straightforward to generalize the present formalism to include them. 
%{\color{red}{PERHAPS: As we will see, the conductance will essentially be proportional to the number of flavors.  }}
%This is a reasonable assumption as these quantum numbers are conserved by dominant scattering mechanisms and thus the conductance will be proportional to the total number of such additional flavors.  COMMENT: The relaxation rate will know about additional quantum numbers. It is not entirely clear that the drift velocity will just be the same. 
\newline
\vspace*{-0.1cm}

We consider the case where the two leads are maintained at different chemical potentials $\mu(\pm\frac{L}{2}){=}\mu_{0}\mp\frac{eV}{2}$ and are much wider than the suspended sample (extending in the third dimension), realizing ideal reflection-free contacts, such that the distribution of inflowing particles is entirely determined by the distribution in the leads.  Here, $V$ is the applied bias voltage and $\mu_{0}$ is the average chemical potential, which coincides essentially with the Fermi energy $\varepsilon_F$ at $T\ll \varepsilon_F$. We assume a finite density of states at the Fermi level, excluding the case where $\mu_{0}$ coincides with the Dirac point. We assume that at the boundaries $x{=}\pm\frac{L}{2}$ the reflection-free contacts with the leads fixes the distribution of the  quasiparticles inflowing from the left and right leads to be given by the equilibrium Fermi-Dirac distribution:
\begin{align}
f\left(\vec{k}\vert_{v_{x}\gtrless0};\,x{=}{\pm}\frac{L}{2}\right){=}f_{0}\left(\vec{k}|_{v_{x}\gtrless0};\,\mu{=}\mu_{0}{\mp}\frac{eV}{2}\right),\label{bc1}
\end{align}
where $f_{0}(\vec{k},\mu)=1/(e^{(E_{\vec{k}}-\mu)/T}+1)$. In the ballistic limit (no interactions nor impurities), the distribution function of the electrons is determined by the respective leads from which the quasiparticles were injected \cite{b22,b23}:
\begin{align}
f_{\rm ball}\left(\vec{k}|_{v_{x}\gtrless0};\,x\lessgtr\pm\frac{L}{2}\right){=}f_{0}\left(\vec{k}|_{v_{x}\gtrless0};\,\mu{=}\mu_{0}\pm\frac{eV}{2}\right).\label{bc2}
\end{align}
In the presence of interactions, in sufficiently long samples, the distribution function is expected to tend to a drifting equilibrium \cite{b2}, i.e.,
\begin{equation}
\begin{split}
f(\vec{k}\,;\,L/2-|x|\ll l_{\mathrm{inel}})=&f_{0}(\vec{k};\mu=\mu_{\mathrm{eq}};v_{d})\\
=&\frac{1}{e^{\frac{E_\vec{k}-k_{x}v_{d}-\mu_{\mathrm{eq}}}{T}}+1}, \label{drifteq}
\end{split}
\end{equation}
in the bulk of the sample, far away from the boundaries. Within linear response the chemical potential will be given by $\mu_{\rm eq}=(\mu_{-L/2}+\mu_{L/2})/2=\mu_{0}$, and the drift velocity $v_{d}$ will be proportional to the bias voltage ${V}$. Below we will employ Boltzmann kinetic theory to calculate $v_{d}$ and the conductance of the {2D} sheet as a function of temperature.
\section{Boltzmann equation}\label{BE}
The quasiclassic Boltzmann kinetic theory can be used to describe transport phenomena if the interactions are sufficiently weak, while quantum interference effects are negligible. %{\color{red}{maybe formulate as a condition on $L_\phi$}} {\color{magenta}{$l_{\phi}\sim l_{\mathrm{inel}}$}}. 
Here we are interested in describing the stationary state which results from a competition between the drift of quasiparticles and binary collisions due to {\em e-e} interactions. These two terms compete as the system relaxes to local equilibrium. The quasiclassical distribution function in the stationary state obeys the equation
\begin{equation}
\begin{split}
v_{x}\partial_{x}f(\vec{k},x)&{=-}\int\!\int\!\int\!\frac{d^{2}k'}{(2\pi)^{2}}\frac{d^{2}p'}{(2\pi)^{2}}\frac{d^{2}p}{(2\pi) ^{2}}\text{W}(\vec{k,k';p,p'})\\
&\Bigg[ f(\vec{k},x)f(\vec{k'},x)[1-f(\vec{p},x)][1-f(\vec{p'},x)]\\
&-f(\vec{p'},x)f(\vec{p},x)[ 1-f(\vec{k},x)][1-f(\vec{k'},x)]\Bigg],\label{BEq1}
\end{split}
\end{equation}
where the left-hand side (LHS) describes the drift. The right-hand side (RHS) is the collision integral for binary collisions, where $W(\vec{k,k';p,p'})$ is the quantum mechanical transition rate for the scattering process $\vec{k,k'}\rightarrow\vec{p,p'}$. Within the Born approximation, the principle of micro-reversibility \cite{b11} [$W(\vec{k,k';p,p'}){=}W(\vec{p,p';k,k'})$] holds for time-reversal-invariant systems, as far as spin-orbit interactions can be neglected \cite{b36}. The transition rate depends on the specific interactions. If screening is strong enough, we can replace the screened Coulomb potential by $u\delta(\vec{r-r'})$ with $u$ characterizing the strength of the short-range interactions. Within the Born approximation, and for particles without additional spinor structure, the rate $W(\vec{p,p';k,k'})$ is given by
\begin{eqnarray}
W(\vec{p,p';k,k'})&=&\frac{4\pi}{\hbar}|u|^{2}\delta(\vec{k+k'-p-p'})\notag \\
&&\times\delta(E_{\vec{k}}+E_{\vec{k'}}-E_{\vec{p}}-E_{\vec{p'}}). \label{deltapotential}
\end{eqnarray}
For weakly screened Coulomb interactions in graphene, a more precise form of the transition rate has been worked out in detail in Ref. ~[\onlinecite{b10}].
\subsection{Parametrization of the nonequilibrium distribution} \label{PNED}
We aim to solve the Boltzmann equation (\ref{BEq1}) in the weak bias regime. For a small bias voltage $V$, the nonequilibrium distribution function $f(\vec{k},x)$ can be linearized around a static equilibrium distribution as $f(\vec{k},x)\equiv f_{0}(\vec{k}){+}\delta f(\vec{k},x)$, where the deviation can be conveniently parametrized as 
\bea
\delta f(\vec{k},x){=}\frac{eV}{T}\,f_{0}(\vec{k})[1{-}f_{0}(\vec{k})]\psi(\vec{k},x). 
\eea
Here, we drop the dependence of $f_{0}$ on the average chemical potential $\mu_{0}$. For small deviations, this is essentially equivalent to a distribution function
\begin{equation}
f(\vec{k};x){=}\frac{1}{e^{\frac{E_{\vec{k}}-\mu-eV\psi(\vec{k},x)}{T}}+1}\,.  \label{distfn}
\end{equation}
The boundary conditions (\ref{bc1}) on the quasi-particles that flow in from the leads are now conveniently expressed as 
\bea
\label{boundaryconditions}
\psi(\vec{k}|_{v_{x}\gtrless0}\,;x{=}\mp L/2) {=}\pm 1 \equiv \psi^{\rm bd}_{\pm}(\vec{k}|_{v_{x}\gtrless0}).
\eea
\subsection{Properties of the collision integral} \label{PCI}
The collision integral has some general properties, which do not depend on the details of the interactions. When expressed in terms of linearized deviation functions $\psi(\vec{k},x)$, the collision term in Eq. (\ref{BEq1}) can be thought of as a linear operator $C$ acting on $\psi$ \cite{b11}: 
\begin{widetext}
\begin{equation}
\begin{split}
(C\psi)(\vec{k},x)&{=}\int \!\int\! \int\! \frac{d^{2}k'}{(2\pi)^{2}}\frac{d^{2}p}{(2\pi)^{2}}\frac{d^{2}p'}{(2\pi) ^{2}}\text{W}(\vec{p,p';k,k'})
f_{0}(\vec{k})f_{0}(\vec{k'})[1-f_{0}(\vec{p})][1-f_{0}(\vec{p'})]\\
&[\psi(\vec{k},x)+\psi(\vec{k'},x)-\psi(\vec{p},x)-\psi(\vec{p'},x)],\label{Cpsi}
\end{split}
\end{equation}
where we have employed the principle of detailed balance in equilibrium,
\begin{equation}
\begin{split}
f_{0}(\vec{k})f_{0}(\vec{k'})[1-f_{0}(\vec{p})][1-f_{0}(\vec{p'})]{=}f_{0}(\vec{p})f_{0}(\vec{p'})[1-f_{0}(\vec{k})][1-f_{0}(\vec{k'})]. \label{detbal}
\end{split}
\end{equation}
\end{widetext}
The operator $C$ is a positive semidefinite Hermitian operator acting on the space of  functions $\psi(\vec{k})$ that are square integrable with respect to the natural inner product 
\begin{equation}
\big<\psi(\vec{k})\rvert\psi(\vec{k})\big>{\equiv}\!\int\!\frac{d^{2}k}{(2\pi)^{2}} \psi^{\ast}(\vec{k})\psi(\vec{k}). \label{innprod}
\end{equation}
This follows immediately from rewriting the  matrix elements of the collision operator in the manifestly positive form
\begin{widetext} 
\begin{align}
\big<\psi(\vec{k})\rvert C\psi(\vec{k})\big>&\equiv\frac{1}{4}\int\!\int \!\int \! \int\frac{d^{2}k}{( 2\pi)^{2}}\frac{d^{2}k'}{(2\pi)^{2}}\frac{d^{2}p}{(2\pi)^{2}}\frac{d^{2}p'}{(2\pi)^{2}}
\text{W}(\vec{p,p';k,k'})[\psi(\vec{k}){+}\psi(\vec{k'})-\psi(\vec{p}){-}\psi(\vec{p'})]^{2} \notag \\
&f_{0}(\vec{k})f_{0}(\vec{k'})[1-f_{0}(\vec{p})][1-f_{0}(\vec{p'})],\label{rate}
\end{align}
\end{widetext}
where the microreversibility has been used again. For normalized eigenfunctions of $C$, these matrix elements are naturally interpreted as collision or relaxation rates associated with that deviation. A central element controlling the collision rate  is the squared quantity 
\bea
\label{scatweight}
\delta\psi^2_{\vec{k}, \vec{k'},\vec{q}} \equiv  [\psi(\vec{k}){+}\psi(\vec{k'})-\psi(\vec{p}){-}\psi(\vec{p'})]^{2},
\eea
which depends on $\vec{k}, \vec{k'}$ and the transferred momentum $\vec{q}= \vec{p}-\vec{k}$ (or $\vec{p}'-\vec{k}$).  We will refer to it as the "scattering weight" of the mode $\psi$  for the  process $(\vec{k}, \vec{k}')\to (\vec{p}, \vec{p}')$. 
\subsection{Modes of the Boltzmann equation}
We can recast the Boltzmann equation as a linear operator equation,  
\bea
v_{x}(\vec{k}) f_{0}(\vec{k})[1{-}f_{0}(\vec{k})] \partial_{x}\psi(\vec{k},x)&\equiv &V_x \partial_{x}\psi(\vec{k},x)\nonumber\\
 &{=}&{-}(C\psi)(\vec{k},x), \label{BEq2}
\eea
where the operator $V_x$ just acts by multiplication with $v_{x}(\vec{k}) f_{0}(\vec{k})[1{-}f_{0}(\vec{k})]$. Formally one can multiply from the left with $V_x^{-1}$ and observe that the operator $V_x^{-1}C$ on the RHS acts on $\vec{k}$ space only while the derivative acts on $x$ only. This suggests to look for special solutions with separated variables, $\psi^{m}(\vec{k},x)\propto\Psi^{m}(\vec{k})\Phi^{m}(x)$, satisfying \cite{b15}

\bea
\partial_{x}\Phi^{m}(x) &{=}&- \alpha_{m}\Phi^{m}(x),\label{BEq3} \\ 
V_x^{-1}C\Psi^{m}(\vec{k}) & {=} &\alpha_{m}\Psi^{m}(\vec{k}). \label{BEq4}
\eea
Equation (\ref{BEq3}) yields $\Phi^{m}(x)\propto\exp[-\alpha_{m}x]$, i.e., a mode decaying exponentially with distance from one of the leads. We might then attempt to seek  the general solution of the Boltzmann equation as a superposition of such solutions. However, this approach is a bit too simple-minded. The reason is that the collision operator is not invertible, because it possesses hydrodynamic zero modes. This property does not allow us to use a simple similarity transform with $C^{-1/2}$ to convert the non-Hermitian eigenvalue equation into a Hermitian problem for which the above approach could then be used; instead, we have to take proper care of the hydrodynamic modes first.

\subsubsection{Hydrodynamic  modes}
The collision operator possesses zero modes because all scattering processes conserve the total energy and the particle number. If additionally scatterings other than {\em e-e} interactions are subdominant and weak, and if umklapp processes can be neglected, the interactions also preserve the total quasimomentum. 
%However, the lattice breaks translation invariance and thus a priori reduces the conservation of momentum to the conservation of quasi-momentum only. However, the conservation of total quasi-momentum $\sum_{\vec{k}\,\mathrm{occ}}\vec{k}$ of the quasi-particle system will still be a good approximation if umklapp processes can be neglected, which is indeed true in present work.\par
Each of these collision invariants imply the existence of a zero mode of the collision operator $C$ \cite{b44}, since a deviation of the distribution function describing a change of the conjugate equilibrium parameter cannot decay. The modes corresponding to number ($N$) and energy ($E$) conservation are
\bea
\label{zeromodes}
\varphi_{N}(\vec{k}) &=& 1,\\
\varphi_{E}(\vec{k}) &=& \varepsilon_{\vec{k}}-\varepsilon_F.
\eea
Note that, as a function of the distance to the Fermi energy, $\varphi_{N}$ is even, whereas  $\varphi_{E}$ is odd.

Quasimomentum conservation (in $x$ direction) in the absence of umklapp scattering furnishes the additional zero mode
\bea
\varphi_{P_x}(\vec{k}) &=& k_x,
\eea
while the zero mode $\varphi_{P_y}(\vec{k}) = k_y$ is irrelevant for our setup, since we assume invariance under inversion of the $y$ coordinate. These zero modes lead to nondecaying modes (with $\alpha=0$) of the Boltzmann equation (\ref{BEq4}). They span the three- or two-dimensional null space ${\cal H}_0$ of $C$, depending on the presence or absence of quasimomentum conservation.
\subsubsection{General solution of the Boltzmann equation}
The general solution of the Boltzmann equation can be constructed by decomposing  the $\vec{k}$ sector of $\psi(\vec{k},x)$ into the null space ${\cal H}_0$ and its orthogonal complement ${\cal H}_\perp$. After some algebra, one finds that the most general $x$-dependent solution of the Boltzmann equation takes the form
\bea
\label{generalsol}
\psi(\vec{k},x)&=&\psi_0(\vec{k})+ (x- C^{-1}V_x)\psi_1(\vec{k}) \\
& +& \sum_{m} w_{m}e^{-\alpha_{m} x} \left(C^{-1/2}+\frac{P_0 V_x^{-1}C^{1/2}}{\alpha_m}\right)\Psi^{m}(\vec{k}),\nonumber
\eea
with $\psi_0(\vec{k})\in {\cal H}_0$ and $\psi_1(\vec{k}) \in {\cal U} \equiv {\cal H}_0 \cap V_x^{-1}{\cal H}_\perp$, and $P_0$ being the orthogonal projector onto $ {\cal H}_0$. The $\Psi^{m}(\vec{k})\in {\cal H}_\perp$ are the eigenvectors with non-vanishing eigenvalues $\alpha_m\neq 0$ of the Hermitian operator 
\bea
A\equiv C^{1/2} V_x^{-1}C^{1/2},
\eea
which acts solely on the subspace ${\cal H}_\perp$, i.e.,
\bea
\label{decaymodes}
C^{1/2} V_x^{-1}C^{1/2}\Psi^{m}(\vec{k})=\alpha_m \Psi^{m}(\vec{k}).
\eea
Note that the operator $A$ has a null space in ${\cal H}_\perp$. It is given by $C^{-1/2}V_x {\cal U}  \subset {\cal H}_\perp$. Together with the orthonormal set of functions $\Psi^{m}$, this null space $C^{-1/2}V_x {\cal U}$ spans all of ${\cal H}_\perp$. 

The coefficients $w_m$ and the functions $\psi_0,\psi_1$ in (\ref{generalsol}) must be determined from the boundary conditions (\ref{boundaryconditions}).
\subsubsection{Weak umklapp scattering}
Note that in the presence of umklapp processes,  $\varphi_{P_x}(\theta)$ is not  an exact zero mode anymore, since a drifting equilibrium will eventually decay due to umklapp processes. Nevertheless, if the collisions that conserve quasimomentum are much stronger than the umklapp processes, there will be a slow "umklapp eigenmode" of $C$, which strongly resembles the drifting equilibrium mode, $\varphi_u \approx  \varphi_{P_x}$. It then has
an eigenvalue $c_u$ far smaller than the next smallest eigenvalue of $C$, which is dominated by much faster momentum-conserving relaxation processes. We will discuss further below how this affects the solution of the boundary value problem.
%, such that on ${\cal H}_\perp$  $C^{-1}$ acts essentially like $C^{-1} \approx c_u^{-1}|\varphi_u\rangel\langel \varphi_u|$.
\subsubsection{Current density}
The total current density carried by an off-equilibrium quasiparticle distribution described by $\psi(\vec{k},x)$ is given by 
\bea
\label{current}
J_{x}&=&e \int  v_{x}(\vec{k})\delta f(\vec{k})\frac{d^{2}k}{(2\pi)^2} = e\frac{eV}{T} \int  V_x \psi(\vec{k})\frac{d^{2}k}{(2\pi)^2}
\notag  \\ % \label{j1}
&=& e \frac{eV}{T}\int  V_x \left[\psi_0(\vec{k})- C^{-1}V_x \psi_1(\vec{k}) \right] \frac{d^{2}k}{(2\pi)^2}.
%&=\frac{W}{2\pi}\frac{e^{2}V k_{F}^{2}}{T}\!\int\limits_{-\infty}^{\infty}\!f(k)(1-f(k))d(\delta\zeta_{k})\!\int\limits_{0}^{2\pi} v_{x}\Psi^{0}(\theta)d\theta \notag\\ \label{j2} 
%&=\frac{W}{2\pi}\frac{e^{2}V k_{F}}{\hbar}h_{1}^{0}\int\limits_{0}^{2\pi}\cos\theta\varphi_{1}(\theta)d\theta  \label{j3}
\eea
Here we have used that the spatially growing or decaying modes of the solution (\ref{generalsol}) do not contribute to current, since with  (\ref{decaymodes}) one can show that
%\bea
%&& \int  V_x C^{-1/2}\Psi^{m} \frac{d^{2}k}{(2\pi)^2}= -\frac{1}{\alpha_m} \int  C^{1/2}\Psi^{m} \frac{d^{2}k}{(2\pi)^2}\nonumber\\
%&&\quad = -\frac{1}{\alpha_m}\langle \varphi_N | C^{1/2}\Psi^{m}\rangle =- \frac{1}{\alpha_m}\langle C^{1/2} \varphi_N | \Psi^{m}\rangle=0,\nonumber
%\eea
\bea
\label{decaycurrent}
&& \int  V_x \left(C^{-1/2}+\frac{P_0 V_x^{-1}C^{1/2}}{\alpha_m}\right)\Psi^{m} \frac{d^{2}k}{(2\pi)^2}\nonumber\\
&& = \frac{1}{\alpha_m} \int  V_x  \left((1-P_0) V_x^{-1}C^{1/2}+P_0 V_x^{-1}C^{1/2}\right)\Psi^{m} \frac{d^{2}k}{(2\pi)^2}\nonumber\\
&&\quad = \frac{1}{\alpha_m}\langle \varphi_N | C^{1/2}\Psi^{m}\rangle = \frac{1}{\alpha_m}\langle C^{1/2} \varphi_N | \Psi^{m}\rangle=0,\nonumber
\eea
using that the zero mode (\ref{zeromodes}) is annihilated by the collision operator, $C^{1/2} \varphi_N=0$. Similarly one shows that the term $x \psi_1$ in $\psi$ does not contribute to the current density. The current density (\ref{current}) is explicitly independent of $x$, as it has to be in a steady state, where the continuity equation requires $dJ_x/dx=0$.  In our setup the conductance per unit sample width ($W$) is simply the current density divided by the applied bias,
\bea
\label{conductance}
\frac{G}{W}=\frac{J_{x}}{V}&=&  \frac{e^2}{T}\int  V_x \left[\psi_0(\vec{k})- C^{-1}V_x \psi_1(\vec{k}) \right] \frac{d^{2}\vec{k}}{(2\pi)^2}.
\eea
The problem of computing the conductance thus boils down to finding the weight of the vectors $\psi_0\in {\cal H}_0$ and $\psi_1\in  {\cal U}$ from the boundary conditions.
\section{Inversion symmetry of the collision operator} \label{ISCO}
Under rather general conditions, the collision operator $C$ is invariant under the inversion of all momenta, $\vec{k}\rightarrow{-}\vec{k}$.  Such an inversion symmetry follows from the principle of microreversibility in the presence of some additional symmetries. At the level of the Born approximation, the inversion symmetry of the collision operator is already ensured by time-reversal symmetry if spin-orbit interactions can be neglected \cite{b36} (such that quasiparticles are well characterized by their quasimomentum $\vec{k}$ only). Beyond the Born approximation, one also needs to invoke space-inversion symmetry of the Hamiltonian to obtain an inversion-symmetric collision operator \cite{b4}. 

Here we further assume that the Hamiltonian is symmetric under the inversion of the $x$ component of the momentum only, $\vec{k}=(k_x,k_y) \to (-k_x,k_y)$, an operation we denote by $I$. This symmetry is of particular interest since the spatial setup of the driven 2D system is symmetric under the inversion $x \to -x$.
Together with inversion symmetry this also implies the symmetry of the Hamiltonian under the reflection $k_y \to -k_y$ in momentum space [corresponding to $\vec{k} \to -I(\vec{k})$]. Together with the symmetry of the spatial setup under the reflection $y \to -y$ we conclude that  the deviation function must satisfy $\psi(\vec{k},x){=}\psi(-I(\vec{k}),x)$. 

The above symmetry assumptions imply that the collision operator is invariant under the reflection $I$, $I C I = C$. The eigenfunctions of $C$  can thus be chosen to have definite parity under $I$. Note also that the velocity operator $v_{x}$, and thus the operator $V_x$, are odd under inversion, $I v_{x} I = -v_x$, $I V_x I = -V_x$. 
From this, and the oddness of the boundary conditions (\ref{boundaryconditions}) under $I$, it follows that the solution to the Boltzmann equation must satisfy
\begin{equation}
\psi(\vec{k},x){=} -\psi(I(\vec{k}),-x).\label{Isymmetry}
\end{equation}
\subsection{Solution of the Boltzmann equation in the presence of inversion symmetry}
In the presence of the symmetry $I$, the exact zero modes $\varphi_N, \varphi_E$ are even under $I$. In contrast, the zero mode $\varphi_{P_x}$, or,  in the presence of umklapp processes, its slowly decaying remnant $ \varphi_u$,  are odd eigenmodes of $C$.

In the absence of umklapp processes, the null space ${\cal H}_0$ is three-dimensional. The requirement that the solution (\ref{generalsol}) obey the symmetry relation (\ref{Isymmetry}) implies, however, that the nondecaying part must be odd under $I$ and thus proportional to $\psi_0 \propto \varphi_{P_x}$.
%\bea
%\psi_0 = w_{P_x} \varphi_{P_x}.
%\eea
The space $\cal U$ is one dimensional and spanned by the (suitably normalized) linear combination of zero modes, 
\bea
\tilde{\varphi}_E = a_N \varphi_N + a_E \varphi_E, 
\eea
such that 
\bea
\label{Uspace}
\langle  \varphi_{P_x} | V_x| \tilde{\varphi}_E \rangle=0. 
\eea
We suggestively denote this mode by $\tilde{\varphi}_E$,
%Thus, $\psi_1 = w_{1}\varphi_1$. 
because at low temperatures, one finds $a_N\approx 0$. Indeed, there the spectrum around $\varepsilon_F$ can be linearized, and the mode $\varphi_{P_x}= k \cos(\theta) \approx k_F \cos(\theta) $, as well as the operator $V_x$, are essentially even as functions of  $\varepsilon_\vec{k} - \varepsilon_F$. Since, in contrast, $\varphi_E$ is odd in this sense, while $\varphi_N$ is even, we see that the orthogonality condition (\ref{Uspace})  essentially selects the mode $\tilde{\varphi}_E \approx \varphi_E$ to span $\cal U$.

The solution of the Boltzmann equation thus takes the form
\bea
\label{fullsol}
\psi(\vec{k},x)&=&w_{P_x} \varphi_{P_x}(\vec{k})+ w_{E} (x- C^{-1}V_x)\tilde{\varphi}_E(\vec{k}) \\
&& +\sum_{m} w_{m}e^{-\alpha_{m} x} \left(C^{-1/2}+\frac{P_0 V_x^{-1}C^{1/2}}{\alpha_m}\right)\Psi^{m}(\vec{k}).\nonumber
\eea
The conductance is then given by the expression (\ref{conductance}) 
\bea
\label{conductance2}
\frac{G}{W}&=&  \frac{e^2}{T} w_{P_x}\int    V_x \varphi_{P_x}(\vec{k}) \frac{d^{2}\vec{k}}{(2\pi)^2}\\
   && - \frac{e^2}{T} w_{E}\int    V_x C^{-1}V_x \tilde{\varphi}_E(\vec{k})  \frac{d^{2}\vec{k}}{(2\pi)^2}.\nonumber
\eea
As we will discuss later, at low temperatures only the first term will have a substantial amplitude. Our task will thus be to determine the coefficient $w_{P_x}$.

\section{Crossover to Ohmic regime with weak umklapp scattering} \label{Umklapp}
It is useful to see how the solution (\ref{fullsol})  arises in the limit of vanishing umklapp scattering from the solution with finite umklapp scattering. If the latter is finite but weak, ${\cal H}_0$ is only spanned by the modes $\varphi_N, \varphi_E$, which are even under $I$, and thus there is no nondecaying part in the solution, i.e., $\psi_0=0$. The space $\cal U$ instead is now two dimensional and coincides with ${\cal H}_0$.

The solution of the Boltzmann equation now reads 
\bea
\label{fullsolwU}
\psi(\vec{k},x)&=&w_{N} (x- C^{-1}V_x){\varphi}_N(\vec{k})\nonumber\\
&&+ w_{E} (x- C^{-1}V_x)\tilde{\varphi}_E(\vec{k})\nonumber \\
%&&+C^{-1/2}\left( \sum_{m} w_{m}e^{-\alpha_{m} x}\Psi^{m}(\vec{k})\right).
& +& \sum_{m} w_{m}e^{-\alpha_{m} x} \left(C^{-1/2}+\frac{P_0 V_x^{-1}C^{1/2}}{\alpha_m}\right)\Psi^{m}(\vec{k}).\nonumber
\eea
Note that the sum over decaying modes remains essentially unchanged. Indeed, in both cases, the $\Psi^m$ span the orthogonal complement of the four-dimensional space ${\cal H}_0 + V_x{\cal U}$, which remains essentially unaffected by turning on weak umklapp scattering. In the limit of weak umklapp scattering, we further have  
\bea
\label{crossover}
w_N(x- C^{-1}V_x){\varphi}_N \approx  w_Nx \phi_N -  w_N\varphi_u \frac{\langle \varphi_{u}| V_x |  \phi_N \rangle}{c_u},\quad\quad
\eea
with the umklapp scattering rate
\bea
c_u \approx \langle \varphi_{u}| C |  \varphi_u \rangle.
\eea
In the limit $c_u \to 0$ and for finite samples smaller than the (diverging) crossover length,
\bea
L_* = \frac{v_F}{c_u},
\eea
$w_N$ is proportional to $c_u$, so that essentially only the second term in (\ref{crossover}) survives. In this limit the conductance $G(L) \approx G_0$ will be nearly independent of $L$, as we will calculate below. For $L \gg L_*$, however, the first term in (\ref{crossover})  dominates (with the coefficient $w_N$ saturating at  $w_N\sim 1/L$) and establishes Ohm's law  with a conductance  that decays as $G(L) \sim G_0 L_*/L = G_0 v_F/ c_u/L$, corresponding to a finite, umklapp-dominated conductivity $\sigma=G_0 v_F/ c_u$.
\section{Reduction to angular variables} \label{RAV} 
So far we have dealt with the conductance problem in full generality, retaining all modes of the Boltzmann equation. However, it turns out that at low temperature many modes will have a negligible weight in the actual solution. It is thus convenient to identify a smaller set of modes of the Boltzmann equation which nevertheless suffices to describe the transport problem accurately. To this end, we recall that in $D=2$ dimensions  forward scattering with small transferred momentum ($\vec{q}\rightarrow 0$) is logarithmically enhanced, as compared to angular relaxation~\cite{b19}. The fast small angle scattering thus rapidly establishes energy relaxation among quasiparticles that move collinearly \cite{b19}. If the dispersion is  linear (like in graphene), such that all quasiparticles have the same velocity, the logarithmic enhancement is present at any temperature and the logarithmic divergence in the collinear scattering cross section  is only cut off by interaction effects \cite{b10}. In a Fermi liquid with quadratic or more generic nonlinear dispersion, the logarithmic enhancement is limited by the nonlinearity. In that case, the logarithmic enhancement only shows at sufficiently low temperatures, $T\ll \varepsilon_F$~\cite{Schuett2011}.
%In what follows, we linearize the spectrum around the Fermi level and work with a constant Fermi velocity $v_F$.
\par
Assuming fast energy relaxation at fixed angles, we concentrate on the remaining angular dependence of $\psi(\vec{k})$ ~\cite{b18,b19,b20} and seek slow modes of the collision operator in the form 
\bea
\label{angular modes1}
\psi(\vec{k})= \psi(\theta),
\eea
which are constant as a function of $|\vec{k}|$, and where  $\theta$ denotes the angle between the wave vector $\vec{k}$ and the $x$ axis, along which the voltage bias is applied. To make this more precise, we observe that since the collision operator has a small expectation value on all functions that are constant as a function of $|\vec{k}|-k_F$ within the thermal window $||\vec{k}|-k_F| \lesssim T/\hbar v_F$, this guarantees that there is a family of slowly relaxing eigenmodes of the collision operator, which are essentially only functions of $\theta$ in the thermally relevant regime $||\vec{k}|-k_F| \lesssim T/\hbar v_F$. The high-energy tails of those modes are likely to deviate from these constants, but we  nevertheless parametrize the modes with the function $\psi(\theta)$ describing their core and restrict the solution of the Boltzmann equation to these modes.\par
Note that there is actually a further family of modes that is not subject to logarithmically enhanced forward scattering, namely,
\bea
\label{angular modes2}
\psi'(\vec{k})= \psi'(\theta) (\varepsilon_\vec{k}-\varepsilon_F),
\eea
which describes an angle-dependent temperature, while the modes (\ref{angular modes1}) can be regarded as describing angle-dependent chemical potentials. In both cases, collinearly moving particles are mutually in equilibrium and thus the logarithmic enhancement of  the forward scattering rate is suppressed. However, the modes (\ref{angular modes2}) are odd with respect to the Fermi wave vector $k_F(\theta)$, whereas the boundary conditions, are even in that sense. Moreover, both  operators $V_x$ and $C$ approximately preserve this even or odd character at low temperatures, where only the vicinity of the Fermi level is relevant. We therefore expect that the energy-odd modes (\ref{angular modes2}) play a negligible role in the solution of our conductance problem and we will neglect them henceforth.\par
To reduce to angular variables, we inject the ansatz  (\ref{angular modes1}) into the Boltzmann equation, multiply the equation from the left with a mode (\ref{angular modes1}) and integrate out the radial variable $k$, which results in the equation
 \bea
B v_x(\theta)  \partial_{x}\psi(\theta,x)&{=}&{-}\int d\theta' C(\theta, \theta') \psi(\theta',x), \label{BEqangular}
\eea
where
\bea
\label{B}
B=\int  f_{0}(\vec{k})[1{-}f_{0}(\vec{k})] k dk, 
\eea
and
\bea
C(\theta, \theta') = \int k dk \int k' dk' C(\vec{k},\vec{k}').
\eea 
Projected onto the slowly relaxing space of modes (\ref{angular modes1}), the collision operator has become a linear operator in the space of angle-dependent functions and can thus be described by a kernel $C(\theta, \theta')$. The operator $V_{x}$ acts  by multiplying a function by $v_{x}(\theta)$, evaluated at  the Fermi surface. For a spherical Fermi surface one has $v_{x}(\theta)= v_F \cos(\theta)$. The scalar product (\ref{innprod}) turns into the standard inner product of functions on the circle $[0,2\pi]$. 

Once we project onto angular variables, the reflection $I$ translates into the mapping of angles $\theta\rightarrow I(\theta) \equiv \pi-\theta$.  It follows from (\ref{Isymmetry}) that the solution of the Boltzmann equation  satisfies, 
\begin{equation}
%(I\psi) \left(\theta,x\right) \equiv 
\psi \left(\theta,x\right) =-\psi\left(I(\theta),-x\right) = -\psi\left(\pi - \theta,-x\right). \label{mode}
\end{equation}
Moreover, the symmetry $k_{y}\rightarrow-k_{y}$ restricts the solution space to even functions under $\theta\rightarrow-\theta$,
\begin{equation}
\psi \left(\theta,x\right) =\psi \left(-\theta,x\right). \label{mode3}
\end{equation}
 \subsection{Zero modes in angular projection}
Upon projection to the angular variables, the zero mode corresponding to particle conservation is a constant, angle-independent deviation, 
\bea
\varphi_{N}(\theta) = {\rm const}
\eea
while the mode related to energy conservation is odd in energy and will thus be neglected.

 The zero mode corresponding to the conservation of quasimomentum in $x$ direction (in the absence of umklapp processes) is given by 
\bea
\varphi_{P_x}(\vec{k})= k_x, 
\eea
or after projection to angular variables, 
\bea
\varphi_{P_x}(\theta)= \cos(\theta).
\eea
It describes an equilibrium state with a finite drift velocity. In a system that conserves total quasimomentum, this mode is anticipated to have a finite amplitude  in the middle of a long sample, while all other deviations from equilibrium have decayed.

As the projection preserves the behavior under inversion, $\varphi_{N}$ is again even under the inversion $I$, while $\varphi_{P_x}$ is odd. We already know from the full solution (\ref{fullsol}) that in the absence of umklapp scattering $\varphi_{N}$ does not enter the solution of our boundary problem.
%Indeed, $\varphi_{1}(\theta)$ satisfies the constraint under inversion $I$, $\varphi_{1}(\vec{k})= - \varphi_{1}(I(\vec{k})) $, or $\varphi_{1}(\theta)= - \varphi_{1}\left(\pi - \theta\right)$. 
\section{Conductance of Systems with Inversion Symmetric Scattering} 
\label{CSISS}
In this section, we calculate temperature corrections to the conductance of 2D systems which obey reflection symmetry and have negligible umklapp scattering. To match the boundary boundary conditions and find the corresponding expansion coefficients for the Boltzmann modes, we need to analyze the decaying modes in more detail.

\subsection{Decaying Modes of the Boltzmann equation}

Considering that the collision operator $C$ is invariant (even) under the reflection $I$, while the velocity operator $V_x$ is odd, it is useful to split the modes $\Psi^{m}$ into their $I$-even and $I$-odd  components,
\bea
\Psi^{m} = \Psi_e^{m}+ \Psi_o^{m}.
\eea
The components obey the equations
\bea
A \Psi_{o}^{m}= C_{e}^{1/2}V_x^{-1}C_{o}^{1/2} \Psi_{o}^{m}\equiv A_{eo}  \Psi_{o}^{m}= -\alpha_m \Psi_{e}^{m} \label{BEq8}, \\
A \Psi_{e}^{m}= C_{o}^{1/2}V_x^{-1}C_{e}^{1/2} \Psi_{e}^{m}\equiv A_{oe}  \Psi_{e}^{m} = -\alpha_m \Psi_{o}^{m} \label{BEq9}. 
%\alpha_{m} V_x|\Psi_{e}^{m}\big\rangle&=&C_{o}|\Psi_{o}^{m}\big\rangle  \label{BEq8}, \\
%\alpha_{m} V_x|\Psi_{o}^{m}\big\rangle&=&C_{e}|\Psi_{e}^{m}\big\rangle \label{BEq9}.
\eea%
Here, we have defined the restrictions of  the collision operator onto the $I$-even and I-odd parity sectors of $\cal H_\perp$, respectively,  $C=C_{o}+C_{e}$,
and the operators
\bea
A_{eo}=C_{e}^{1/2}V_x^{-1}C_{o}^{1/2}=A_{oe}^\dagger.
\eea
As we will discuss in Sec.~\ref{RM} below, the eigenvalues of $C_{e,o}$ scale parametrically differently with temperature in the two sectors, the even modes relaxing much faster than the odd ones cf. Eq.~(\ref{ratio}) below.
%, namely,
%\bea
%XXX
%\eea
This will allow us to derive general properties of the Boltzmann modes and the solution to our transport problem without the need to specify further microscopic details of the collision operator.

%We expand the modes $\Psi^{m}(\theta)$ of the Boltzmann equation in this basis as $\Psi^{m}(\theta)={\sum\limits_{n\geq0}^{\infty}}h_{n}^{m}\varphi_{n}(\theta)$ with expansion coefficients $h_{n}^{m}$. By construction, the collision operator acts diagonally, while the drift operator mixes the basis states. %As opposed to $C$, $v_{x}$ is neither Hermitian nor necessarily positive definite. 
%The complete solution $\psi(\theta,x)$ of the Boltzmann equation consists of conserved, spatially non-decaying modes and exponentially decaying or increasing modes, corresponding to eigenvalues $\alpha_{0}=0$ and $\alpha_{m>0}\neq0$ respectively. These modes are discussed in detail in the next sections.

Combining Eqs.~(\ref{BEq8}) and \ref{BEq9}), we obtain the eigenvalue equation
\bea
\label{BEq10} 
A_{oe}A_{eo}{\Psi}_{o}^{m} = A_{eo}^\dagger A_{eo}{\Psi}_{o}^{m}= \alpha_{m}^{2} \Psi_{o}^{m}.
%\left[ C_{o}^{-\frac{1}{2}} V_xC_{e}^{-\frac{1}{2}}\right] \left[  C_{o}^{-\frac{1}{2}}V_xC_{e}^{-\frac{1}{2}}\right]  ^{\text{T}}|\bar{\Psi}_{o}^{m}\big>&=\frac{1}{|\alpha_{m}|^{2}}|\bar{\Psi}_{o}^{m}\big>. \label{BEq11}
\eea
As we already mentioned, the subspace $V_x{\cal U}$ spans the (odd) zero modes of this equation. On its orthogonal complement in $\cal H_\perp$, we expect the operator $A_{eo}^\dagger A_{eo}$ to act as a positive definite operator. Let us  label its eigenmodes with a new index $n$.

Note that the full modes $\Psi^m$ come in pairs: Every eigenmode $\Psi^n_o$ of (\ref{BEq10}) with positive eigenvalue $\alpha_n$ gives rise to two inversion-related modes,
\bea
\Psi^n_{\pm}= \Psi^n_o\mp  \frac{1}{\alpha_n} A_{eo}\Psi^n_o\equiv  \Psi^n_o \pm \Psi^n_e,
\eea
which differ in the relative sign of the even and odd components. We adopt the convention that $\alpha_n>0$ from now on. Note that $\alpha_n$ is the spatial rate of decay of the modes $ \Psi^n_{\pm}$ away from the leads. The smallest positive eigenvalue can thus be related to an inelastic relaxation length scale,  
\bea 
\label{linel1}
\ell_{\rm inel } = \frac{1}{{\rm min}_n(\alpha_n)}.
\eea
Note that the positive eigenvalues $\alpha_n$ scale like $\sqrt{c_o c_e}$ where $c_{o,e}$ are typical eigenvalues of the collision operator in the odd and even sectors, respectively.

The symmetry under inversion $I$ implies that the solution (\ref{fullsol}) of the Boltzmann equation is constrained to take the form 
\begin{widetext}
\bea
\label{fullsolsym}
&& \psi(\vec{k},x)=w_{P_x} \varphi_{P_x}(\vec{k})+ w_{E} (x- C^{-1}V_x)\tilde{\varphi}_E(\vec{k}) \\
&&  \quad +\sum_{n} w_{n} e^{-\frac{\alpha_{n} L}{2}} \left( 2\cosh(\alpha_n x)\left(C^{-1/2}_o\Psi^{n}_o(\vec{k}) +\frac{P_0V_x^{-1}C_e^{1/2}}{\alpha_n} \Psi^{n}_e(\vec{k})\right) 
% \right.
%\nonumber \\
%&&  \left. 
- 2\sinh(\alpha_n x)\left(C^{-1/2}_e \Psi^{n}_e(\vec{k})+\frac{P_0V_x^{-1}C_o^{1/2}}{\alpha_n}\Psi^{n}_o(\vec{k})\right) \right),\nonumber
\eea
where the weights $w_{P_x}, w_{E}, w_n$ are to be determined from the boundary conditions (\ref{boundaryconditions}).
\end{widetext}

The even and odd components of $\Psi^n_{\pm}$ are generically of comparable norm. However, in the solution (\ref{fullsolsym}) they enter as $C_{o,e}^{-1/2} \Psi^n_{o,e}$. Since typical eigenvalues of $C_e$ are parametrically  bigger than those of $C_o$, the odd components dominate in the mode expansion, while the even components are suppressed by a factor $O(({c_{o}/c_{e}})^{1/2})\ll 1$, where $c_{o,e}$ are typical eigenvalues of $C$ in the odd and even sectors, respectively. We will make use of this feature below to solve for the boundary conditions.

Before doing so, we briefly discuss the collision rates and the nature of the dominant processes involved in the even and odd sectors, respectively.

\subsection{Relaxation rates from collisions} \label{RM}
The relaxation rate of eigenmodes $\Phi^m$ of the collision operator are simply given by its eigenvalue, which one can express as
\bea
c_{m} = \frac{\big<\Phi^m(\vec{k})\rvert C\Phi^m(\vec{k})\big>}{\big<\Phi^m(\vec{k})\rvert \Phi^m(\vec{k})\big>}
\eea
and the expression (\ref{rate}) can be used to evaluate the numerator. The dependence on temperature can be estimated as the product of two factors: (i) the phase space of a certain type of kinetically allowed  scattering processes and (ii) the associated scattering weight {{$\delta\psi^2_{\vec{k}, \vec{k'},\vec{q}}$}}, cf. (\ref{scatweight}). This analysis is carried out in detail in the appendix. %~\ref{Relrates}.
It turns out that for the relevant slow modes that have essentially only an angular dependence, head-on collisions are the most efficient channel for relaxation. They come with a phase-space volume that scales like $T^2$. However,  modes that are odd or even in momentum space, respectively, behave very differently with respect to the scattering weight. In even modes, the  contributions to the scattering weight from counterpropagating quasiparticles add up, while they essentially cancel in odd modes. The latter suppresses the relaxation rate of odd modes  by an additional factor of $T^2$ (up to a logarithmic enhancement) as compared to that of the much faster relaxing even modes \cite{odd}. In summary, we find the relaxation rates
%So the head on collision dominates the relaxation of angular modes as comprehended in Sec. \ref{RAV}. \par
%After determining the scattering weight, we are in a position to evaluate their relaxation rates for head on collision process using Eq. (\ref{rate}). For time being, we are interested only in temperature dependence of these rates, which can be determined by counting the powers of $T$ that have contribution from integration over momenta of transition rate W$(\vec{k,k';p,p'})$ and scattering weight $\delta\Psi$ as done in appendix \ref{TDRR}. In particular, the relaxation rates of modes are observed to have the following temperature dependence (in units of $\frac{\varepsilon_F}{\hbar}$); $
\bea
\label{ceo}
c_{e} &\sim&\left(\frac{T}{\varepsilon_F}\right)^{2},\\
c_{o} &\sim&\left(\frac{T}{\varepsilon_F}\right)^{4}\log\frac{\varepsilon_F}{T},
\eea
 as long as head-on collisions are the dominant relaxation channel. The logarithmic enhancement in the relaxation rate for odd modes $c_{o}$ is due to singular small-angle scattering in $2D$ \cite{b4,b12,b13,b14,b50}. Gurzhi {\em et al.} reported similar rates in Ref.~[\onlinecite{b21}], however, without pointing out the logarithmic factor in $c_{o}$. 
%when it changes form $O(1)$ to $O\big(\frac{T}{\varepsilon_F}\big)$. 
The ratio of collision rates in the odd and even sectors thus provides a small parameter
\begin{equation}
\frac{c_{o}}{c_{e}}\sim\left(\frac{T}{\varepsilon_F}\right)^{2}\log\frac{\varepsilon_F}{T},\label{ratio}
\end{equation}
while the inelastic length scale (\ref{linel1}) grows as  
\bea
\label{inel}
\ell_{\rm inel}(T)\sim\frac{1}{k_{F}}\left(\frac{\varepsilon_F}{T}\right)^{3} \left({\log\frac{\varepsilon_F}{T}}\right)^{-1/2},
\eea
with decreasing temperature.

%Coming back to expression (\ref{dm13}), the total solution of Boltzmann equation will mainly comprise linear combination of odd modes while the even modes only add temperature dependent corrections $\sim\frac{T}{\varepsilon_F}\sqrt{\log\frac{\varepsilon_F}{T}}$ to total solution so that expression (\ref{dm13}) gives
%\begin{equation}
%\left\Vert\Psi^{m}\right\Vert\sim\left[1+\frac{T}{\varepsilon_F}\sqrt{\log\frac{\varepsilon_F}{T}}\right]\left\Vert\Psi_{o}^{m}\right\Vert . \label{dm14}
%\end{equation}

%In the next section we calculate the current density by applying the boundary condition on distribution function of incoming particles. We will see that for infinite sample, the boundary condition reduces to influx from left lead only.
\subsection{Conductance} \label{CD}
The general symmetry-obeying solution (\ref{fullsolsym}) of the Boltzmann equation must match the boundary condition $\psi(\vec{k},x=-L/2)=\psi^{\rm bd}_{+}(\vec{k})=1$, for all $k_x>0$ (where for definiteness we now assume $k_x$ and $v_x$ to have the same sign); cf. Eq.~(\ref{boundaryconditions}). This condition concerns only inflowing wave vectors, i.e., only half of all $\vec{k}$ space, which renders the matching non-trivial. Explicitly, we have
\bea
\label{bdc}
&&\psi^{\rm bd}_{+}(\vec{k}| k_x>0) =w_{P_x} \varphi_{P_x}(\vec{k}) %+ w_{E} (-L/2+ C^{-1}V_x)\tilde{\varphi}_E(\vec{k}) 
\\
&&\quad +\sum_{n} w_{n} (1+e^{-\alpha_{n} L}) \left(C^{-1/2}_o \Psi^{n}_o(\vec{k}) +\frac{P_0V_x^{-1}C_e^{1/2}}{\alpha_n}\Psi^{n}_e(\vec{k})\right)  \nonumber \\
&& \quad +\sum_{n} w_n (1-e^{-\alpha_{n} L}) \left(C^{-1/2}_e \Psi^{n}_e(\vec{k}) +\frac{P_0V_x^{-1}C_o^{1/2}}{\alpha_n}\Psi^{n}_o(\vec{k})\right).\nonumber
\eea
Here we have dropped the contribution $\propto w_E$, since it is associated with modes that are essentially odd as a function of $\varepsilon_{\vec{k}}-\varepsilon_F$. Since the boundary condition and the dominating modes are basically even in  $\varepsilon_{\vec{k}}-\varepsilon_F$, we expect that $w_E$ is at most $O(T/\varepsilon_F)$ and thus contributes  corrections to the conductance that are smaller than the leading ones we derive below. We therefore drop those terms from now on.

We can make progress by observing that the RHS of Eq.~(\ref{bdc}) is nearly an odd function of $\vec{k}$, the even components being smaller by a suppression factor $O((c_o/c_e)^{1/2})$. We can make use of this fact to find the coefficients $w_n$ and $w_{P_x}$ in the form of an expansion in  $(c_o/c_e)^{1/2}$.

For a function $g(\vec{k})$ defined only on the half space $k_x>0$, let us define the odd (under inversion $I$) function (for all $\vec{k}$)
\bea
\overline{g}(\vec{k}) := g(\vec{k}) \Theta(k_x) - g(I(\vec{k})) \Theta(-k_x),
\eea
where $\Theta$ is the Heaviside function.

Let us now apply this operation to both sides of (\ref{bdc}), observing that odd functions are unchanged under the above operation, $\overline{\Psi^{n}_o}= \Psi^{n}_o$ for all $\vec{k}$,
\bea
\label{expension}
&&\overline{\psi^{\rm bd}_{+}}(\vec{k}) -\lambda \sum_{n} w_n (1-e^{-\alpha_{n} L}) \left(\frac{c_e}{c_o}\right)^{1/2}\nonumber\\
&& \quad \times \overline{\left(C^{-1/2}_e \Psi^{n}_e +\frac{P_0V_x^{-1}C_o^{1/2}}{\alpha_n}\Psi^{n}_o(\vec{k})\right)},\nonumber\\
&&=w_{P_x} \varphi_{P_x}(\vec{k}) %+ w_{E} (-L/2+ C^{-1}V_x)\tilde{\varphi}_E(\vec{k}) 
 +\sum_{n} w_{n} (1+e^{-\alpha_{n} L}) \nonumber\\
&& \quad \times \left(C^{-1/2}_o \Psi^{n}_o(\vec{k}) +\frac{P_0V_x^{-1}C_e^{1/2}}{\alpha_n}\Psi^{n}_e(\vec{k})\right).
% C^{-1/2}_o\Psi^{n}_o(\vec{k}). 
\eea
We have moved the parametrically small contribution of the even part to the LHS and multiplied the terms under its sum by a factor $\left(c_e/c_o\right)^{1/2}$, to make them of the same order as the odd terms on the RHS. This is compensated for by the prefactor $\lambda$, which is eventually to be set to 
\bea
\lambda=\left(\frac{c_o}{c_e}\right)^{1/2}, 
\eea
but now serves us as a small expansion parameter. For definiteness, we define the rates $c_{o,e}$ as the smallest positive eigenvalues of $C_{o,e}$, respectively.

We now expand the coefficients in (\ref{bdc}) as a formal power series in $\lambda$,
\bea
w_{P_x}&=& \sum_{\ell=0}^\infty w_{P_x}^{(\ell)}\lambda^\ell,\\
w_{n}&=& c_o^{1/2}\sum_{\ell=0}^\infty w_{n}^{(\ell)}\lambda^\ell.
\eea 
The successive steps in perturbation theory amount to solving equations of the form
\bea
\label{oddfunctions}
w_{P_x}^{(\ell)} \varphi_{P_x} \nonumber%+ w_{E} (-L/2+ C^{-1}V_x)\tilde{\varphi}_E(\vec{k}) 
 &+&\sum_{n} w_{n}^{(\ell)} (1+e^{-\alpha_{n} L}) c_o^{1/2} \\
 &&\quad \times \left(C^{-1/2}_o \Psi^{n}_o(\vec{k}) +\frac{P_0V_x^{-1}C_e^{1/2}}{\alpha_n}\Psi^{n}_e(\vec{k})\right).  \notag\\
  &=& \overline{g}^{(\ell)},
\eea
with 
\bea
\label{gell}
\overline{g}^{(\ell=0)}&=&\overline{\psi^{\rm bd}_{+}},\\
\overline{g}^{(\ell>0)}&=&- \sum_{n} w_n^{(\ell-1)} (1-e^{-\alpha_{n} L}) c_e^{1/2}\nonumber\\
\label{gell2}
&& \times\overline{\left(C^{-1/2}_e \Psi^{n}_e + \frac{P_0V_x^{-1}C_o^{1/2}}{\alpha_n}\Psi^{n}_o(\vec{k})\right)}.
%\overline{C^{-1/2}_e\Psi^{n}_e}.\nonumber
\eea

Equation~(\ref{oddfunctions}) is an  equation in the space of odd functions. We can solve for the expansion coefficients by exploiting the orthogonality properties of the Boltzmann modes. We assume that  the eigenvalue problem (\ref{BEq10}) has been solved, i.e., that all modes and decay rates $\alpha_n$ of the Boltzmann equation have been determined.
 
Acting with $c_o^{-1/2} C_o^{1/2}$ on both sides of (\ref{oddfunctions}), one obtains,
\bea
\label{oddfunctions2}
\sum_{n} w_{n}^{(\ell)} (1+e^{-\alpha_{n} L}) {\Psi^{n}_o} = c_o^{-1/2}C_o^{1/2}\overline{g}^{(\ell)}.
\eea
Now one can use the orthonormality of the $\Psi^{n}_o$ [which are eigenfunctions of the Hermitian operator (\ref{BEq10})] to find
\bea
w_{n}^{(\ell)} = \frac{1}{(1+e^{-\alpha_{n} L})}\langle \Psi^{n}_o | c_o^{-1/2}C_o^{1/2}\overline{g}^{(\ell)}\rangle.
 \eea
To determine $w_{P_x}$, we multiply (\ref{oddfunctions}) by the velocity operator $V_x$ and integrate over all $\vec{k}$. Since this operation applied to the total deviation function $\psi$ actually computes the current density, it projects out all decaying or increasing modes. Indeed, as shown earlier, $\int d^2k V_x C^{-1/2}_o\Psi^{n}_o=0$. We thus find
 \bea
 \label{current2}
 w_{P_x} \int  V_x \varphi_{P_x} \frac{d^{2}\vec{k}}{(2\pi)^2} =  \sum_{\ell=0}^{\infty} \left(\frac{c_o}{c_e}\right)^{\ell/2} \int  V_x \overline{g}^{(\ell)} \frac{d^{2}\vec{k}}{(2\pi)^2}.
 \eea
Comparing with (\ref{conductance2}), the LHS of (\ref{current2}) (up to neglected subleading contributions $\sim w_E$ due to the zero mode that is odd in energy) is  seen to yield $G/W (T/e^2)$, that is, the sought conductance. The first term on the RHS of (\ref{current2}) can be shown to  yield, up to simple prefactors, the ballistic Landauer-B\"uttiker conductance~\cite{b7,b16,b17},
 \bea
 &&\int  V_x \overline{g}^{(0)} (\vec{k})\frac{d^{2}\vec{k}}{(2\pi)^2} =\int  V_x  \overline{\psi^{\rm bd}_{+}}(\vec{k}) \frac{d^{2}\vec{k}}{(2\pi)^2}\nonumber\\
  &&\quad=  -\int_{k_x>0}  v_x \frac{df}{d\varepsilon_\vec{k}}  \frac{d^{2}\vec{k}}{(2\pi)^2}+ \int_{k_x<0}  v_x \frac{df}{d\varepsilon_\vec{k}}  \frac{d^{2}\vec{k}}{(2\pi)^2} \nonumber\\
  &&\quad= \frac{T}{e^2}\frac{G_{\rm ball}}{W}. 
 \eea
Naturally, the ballistic conductance should  result in the limit of very short samples ($L\ll \ell_{\rm inel}$) where scattering is irrelevant. Indeed, we see from (\ref{gell2}) that the corrections $\overline{g}^{(\ell>0)}$ tend to zero in this limit, and we thus correctly capture the noninteracting, ballistic limit from the leading term $\ell=0$.

Much less trivial, however, is the statement we obtain in the opposite limit of samples of length $L\gg \ell_{\rm inel} $ (but still much smaller than the scale where umklapp becomes relevant, $L\ll L_*$). Namely, we find that collisions lead to nontrivial corrections of the conductance which scale as
\bea
\label{corrtoballistic}
&&\frac{G-G_{\rm ball}}{W} = \frac{e^2}{T} \sum_{\ell=1}^{\infty} \left(\frac{c_o}{c_e}\right)^{\ell/2} \int  V_x \overline{g}^{(\ell)} \frac{d^{2}\vec{k}}{(2\pi)^2}\nonumber\\
&&\quad = O\left(\left(\frac{c_o}{c_e}\right)^{1/2}\right)= O\left(\frac{T}{\varepsilon_F} [\log(\varepsilon_F/T)]^{1/2}\right).
\eea
The surprising aspect of this result is that, in the limit $L\gg \ell_{\rm inel} $, many collisions take place as the current traverses the sample, and thus there is {\em a priori} no reason to expect a conductance close to the noninteracting, ballistic value. Nevertheless, we find here that in the presence of inversion symmetry, the corrections to the ballistic result are small and tend to zero with $T\to 0$, even if the sample length is kept much larger than the (diverging) inelastic relaxation scale $\ell_{\rm inel}$. The reason for this rather unexpected behavior is not simply the  inefficiency of collisions at low $T$ in absolute terms, but rather the parametrically large difference in the relaxation rates of even and odd modes of quasiparticle excitations. The basic mechanism behind this phenomenon is the following: A spatial gradient in an odd distortion mode in momentum space generates an even component due to the drift of the quasiparticles. However, this even part relaxes very quickly under head-on collisions, before further drift could develop a substantial odd component that would diminish the current carrying and nondecaying odd mode. Under these circumstances, the quasiparticle distribution  remains very close to being odd in momentum space, and the amplitude of the current-carrying mode remains close to its weight in the noninteracting limit. Nevertheless, there is a finite, if small amount of backscattering due to {\em e-e} collisions, which is the correction term we have computed above. It is natural to expect that in general the correction on the RHS of Eq.~(\ref{corrtoballistic}) is negative.

Note that thermal corrections to the zero-temperature conductance arise also from the  thermal smearing of the Fermi surface. This effect is already present in the Landauer-B\"uttiker conductance. However, it is usually  weak and scales as $T^2$.

\begin{figure}[h]
\begin{center}
\includegraphics[width=\linewidth]{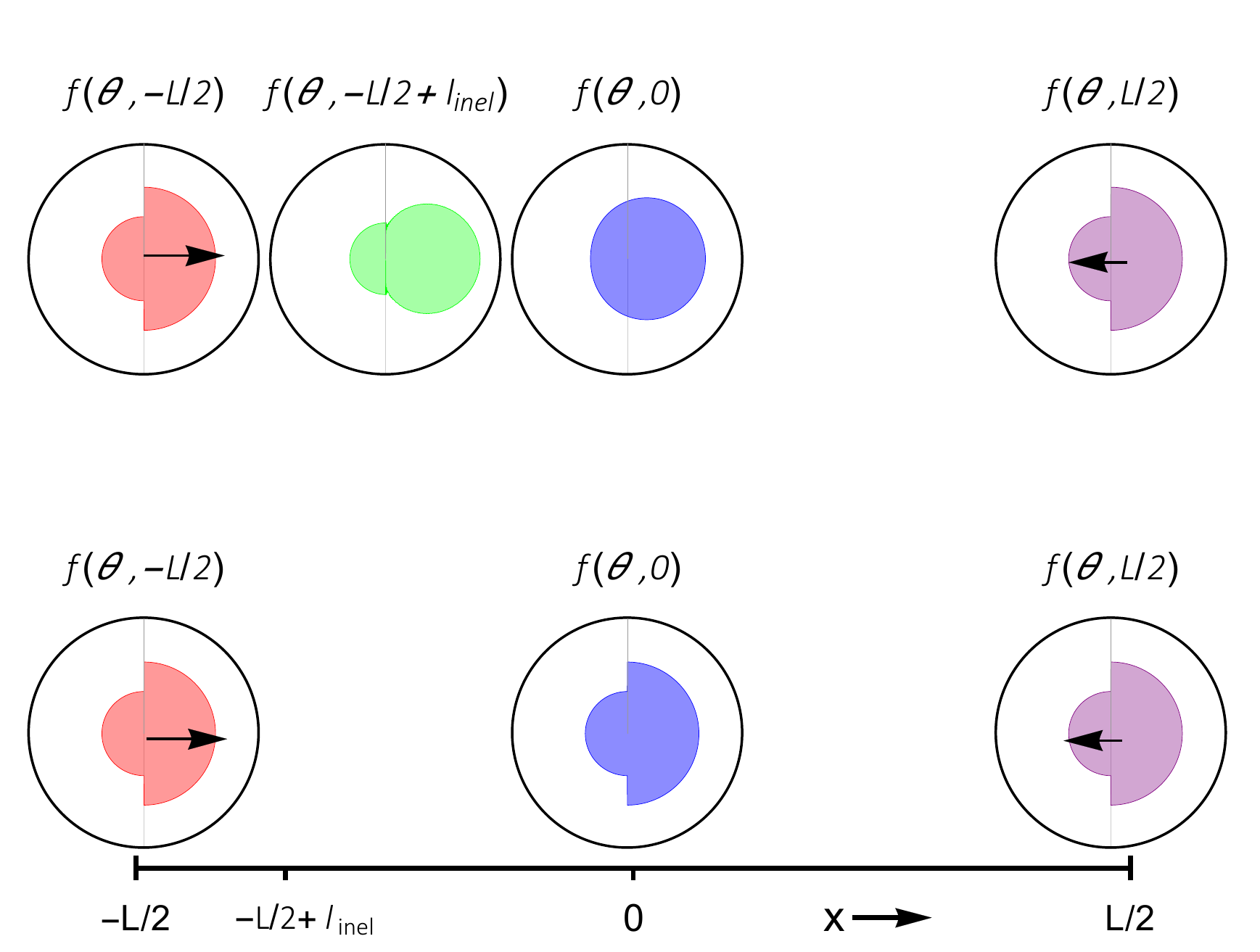} \vspace*{-1cm} 
\end{center}
\vspace*{0 cm}
\caption{The upper row illustrates the distribution of  quasiparticles flowing in from two leads. Close to the leads, the incoming particles retain the distribution of the leads, and the full distribution resembles the one observed in noninteracting ballistic transport, as shown in the bottom row for comparison.
 However, within a distance $\ell_{\rm inel}$ from the leads inelastic {\em e-e} scattering relaxes the distribution to a drifting equilibrium, i.e., a displaced Fermi sphere. Despite the substantial  difference in the quasiparticle distribution, the conductances in the two cases only differ by a term scaling as $T/\varepsilon_F [\log(\varepsilon_F/T)]^{1/2}$ at low temperature.}
\label{fig:2}
\end{figure}

In the upper row of Fig. \ref{fig:2}, we illustrate how the steady-state quasiparticle distribution varies in space. After a distance of the order of $\ell_{\mathrm{inel}}$ from the leads, the drifting equilibrium state is reached, with a displaced Fermi surface. This is to be compared to the noninteracting ballistic case, where the quasiparticle distribution consists of two half Fermi surfaces for left and right movers, respectively, which remains constant throughout the sample. 
%The distribution functions for ballistic and fully equilibrated system are also shown to demonstrate the effect of interactions on transport in an otherwise ballistic sample.
\vspace*{-0.25cm}

\section{Rotationally invariant case} \label{RI}
Let us now consider the simple case in which the Fermi surface and the collision operator $C$ are rotationally invariant. In this case, we can follow through the above procedure in a  quantitative manner, since it allows us to have explicit expressions for the modes, the relaxation rates of the collision operator, and eventually a numerical value for the conductance. To have a concrete system in mind, one may think of a lightly doped graphene sheet with a spherically symmetric Fermi surface, whereby we restrict scattering to a single Dirac cone. In reality this  is often a good approximation, since {\em e-e} scattering between the two Dirac cones is comparatively weak. For the interactions, we will, however, consider the simplest possible short-range interaction, characterized by a momentum-independent transition rate $W(\vec{k,k';p,p'})\to u^2$, where $u$ characterizes the strength of the interaction. This crude approximation neglects effects due to the pseudospinor structure in the collision kernel. Nevertheless, this approximate modeling  gives us reasonable estimates of the inelastic length, and the conductance corrections, which would be interesting to confront with experiments on interaction-dominated (and viscous) flow  reported recently in this type of system \cite{b44,b61,b45,b46,b56,b61,Moll1061}.

For a rotationally invariant collision operator, the angular dependence of its eigenmodes $\Phi_\ell$ is simply an angular harmonic, $\cos(n_\ell \theta)$ (due to the  symmetry under $y \to -y$, we can restrict to even functions of $\theta$),
\bea
\Phi_\ell(\vec{k}) = \phi_\ell(k) \cos(n_\ell \theta).
\eea
For a given angular harmonic $n$, most modes will relax fast due to logarithmically enhanced forward scattering. Only two eigenmodes are expected to remain logarithmically slower. Those will behave as $\phi_\ell(k) \sim {\rm const.}$ or $\sim k-k_F$ within the thermal window  $|k-k_F| \lesssim T/\hbar v_F$, so as to suppress the forward scattering divergence. As argued previously, at low $T$ we may restrict ourselves to the first kind of mode, which is even in $k-k_F$ and thus most relevant to solve our boundary problem. We hence work within a restricted space of eigenfunctions of $C$ which we label solely by the angular harmonic $n$, 
\bea
\label{Phin}
&& \Phi_n(\vec{k}) = \phi_n(k) \cos(n \theta),  \\
&& {\rm with  } \quad \phi_n(k) = 1 \quad {\rm for } |k-k_F| \lesssim T/\hbar v_F.\nonumber
\eea 

In what follows we will  approximate $\phi_n(k) = 1$ for all $k$, since the contributions to integrals over $\phi_n(k)$ from outside the thermal window are small anyway. 

Injecting the ansatz $\psi({\theta,x}) = \sum_n a_n(x) \Phi_n(\theta)$ into the Boltzmann equation (\ref{B}) and projecting it onto the mode $\Phi_{m}(\vec{k})=\Phi_m(\theta)= \cos(m \theta)$, we find the projected Boltzmann equation
\bea
&& B \int d\theta  \cos(m \theta) v_F\cos(\theta) \sum_n \partial_x a_n(x) \cos(n\theta) \\
%&& \quad = \frac{\pi}{2}v_F B \left[(\partial_x a_{m+1}(x)+ \partial_x a_{m-1}(x))(1-\delta_{m,0}) + 2\delta_{m,0} \partial_x a_{1}(x) \right]\nonumber\\
&& \quad = - \tilde{c}_m a_m(x). 
\eea
where $B$ was defined in (\ref{B}) and
\bea
\tilde{c}_m = \left\langle\Phi_{m}(\vec{k})|C\Phi_{m}(\vec{k})\right\rangle,
\eea
are the eigenvalues of the collision operator, which are explicitly given by Eq. (\ref{rate}).
%Once we project onto angular variables,  the velocity operator $V_x$  boils down  to multiplication by $B v_F \cos(\theta)$, where $B$ was defined in (\ref{B}).
%The collision operator essentially takes the form of an angular kernel 
%\bea
%C\left( \theta,\theta'\right) ={\sum\limits_{n=0}^{\infty}}c_{n}'\cos n(\theta-\theta').
%\eea
%where the $c_n$ are proportional (up to a normalization) to the eigenvalues corresponding to the eigenmodes (\ref{Phin}). 
The zero modes of the collision operator are represented by the lowest two angular harmonics, $n=0,1$, associated with charge and momentum conservation, respectively, and thus
$\tilde{c}_0=\tilde{c}_1=0$.
The components of the projected equation then read 
\bea
\label{rotinvBEq}
\partial_x a_{1}(x)&=& 0,\\
\partial_x a_{m+1}(x)+ \partial_x a_{m-1}(x) &=& - c'_m a_m(x),
\eea
where 
\bea
\label{ctilde}
c'_m= \frac{2\tilde{c}_m}{\pi B v_F},
\eea
is the inverse of the relaxation length scale.
%The remaining $\tilde{c}_n$  can be expressed as 
%\bea
%\label{rateApp}
%\tilde{c}_n =\left\langle\Phi_{n}(\vec{k})|C\Phi_{n}(\vec{k})\right\rangle 
%\eea
%whereby the matrix elements of $C$ are obtained from Eq. (\ref{rate}). To calculate the exact eigenvalues $c_n$
%we would need to know the precise shape of the functions $\phi_n(k)$. Unfortunately, we only know that they are essentially constant in the thermal window $|k-k_F| \lesssim T/\hbar v_F$, while we do not know how  they fall off far from the Fermi level, even though it is very likely that they do so exponentially. We can nevertheless estimate the  eigenvalues $c_n$ by evaluating the denominator in (\ref{rateApp}) using a reasonable guess for the high energy tail of  $\phi_n(k)$, such as
%\bea
%\phi_n(k) = 4 f(\hbar v_F (k-k_F))[1- f(\hbar v_F (k-k_F)],
%\eea
%while plugging in the  constant $\phi_n(k=k_F)=1$ when evaluating the numerator of (\ref {rate}), since the full collision kernel cuts off the high energy tail anyway. 

\subsection{Conductance of a graphene sheet} \label{CGS}

For a short-range interactions as described above, we have calculated the "rates" $c'_m$, as outlined in the appendix. The result is 
\begin{align}
c'_{n>1, {\rm odd}}&{=}0.55 (u\rho)^{2} n^{4} \frac{\varepsilon_F}{\hbar}\left(\frac{T}{\varepsilon_F}\right)^{4}\log\frac{\varepsilon_F}{T}, \label{odd}\\
c'_{n>1, {\rm even}}&{=}0.20 (u\rho)^{2} \frac{\varepsilon_F}{\hbar}\left(\frac{T}{\varepsilon_F}\right)^{2},\label{even}
\end{align}
where $u\rho$ is the dimensionless interaction constant, $\rho$ being the density of states at the Fermi level. It reflects explicitly the scaling (\ref{ceo}) anticipated earlier.

%The numerical prefactors in (\ref{odd},\ref{even}) are subject to the uncertainties related to the tail of the shape function $\phi_n(k)$. However,  modifying the shape function for all modes, essentially multiplies all rates by the same factor.  
%Since the final result for the conductance is sensitive only to the ratio of relaxation rates, but not to their absolute values, we expect that the above approximation affects the conductance only by little. 

The projected Boltzmann equation can be solved for decaying and/or increasing Boltzmann modes by truncating the above equations and restricting the modes of the collision operator to angular harmonics $m$ below some cutoff $N$. After determining all Boltzmann modes, we solved the boundary value problem and evaluated the conductance as described in previous sections.

The dimensionless conductance is given by Sharvin's formula (Sharvin contact resistance)  for graphene \cite{b61}
\bea
\frac{G_{\rm ball}}{e^2/ \hbar}=\frac{4k_{F}}{2\pi},
\eea
with $k_{F}=2\sqrt{\pi n}$, $n$ being the carrier density. We have not incorporated spin and valley degeneracy which would simply result in multiplication by a factor of 4.

%The ballistic conductance 
%in Eq. (\ref{j5}) simplifies to 
%\begin{align}
%%h_{1}^{0}&=\frac{4}{\pi}-O\left(\frac{T}{\varepsilon_F}\sqrt{\log\frac{\varepsilon_F}{T}}\right) \notag\\ %\label{G1}
%%J_{x}&=\frac{W}{2\pi}\frac{e^{2}Vk_{F}}{\hbar}h_{1}^{0}\pi \notag \\  %\label{G2}
%G&=\frac{k_{F}}{2\pi}\pi\left[\frac{4}{\pi}-O\left(\frac{T}{\varepsilon_F}\sqrt{\log\frac{\varepsilon_F}{T}}\right)\right] .\label{G3}
%\end{align}
%Till now we have neglected the $n$ dependence in the relaxation rates of modes but they do depend on $n$ for a real system, which in our case is a graphene sheet. The relaxation rates of modes responsible for equilibration in a long ($L\gg l_{inel}$ so that the system equilibrates fully) and wide ($W\rightarrow\infty$ so that boundary effects can be neglected) graphene sheet, considering head on collisions as the dominant scattering process are 
%\begin{align}
%c_{2n+1}&{=}0.55(2n+1)^{4}|u|^{2}\rho^{2}(\varepsilon_F)\frac{\varepsilon_F}{\hbar}\left(\frac{T}{\varepsilon_F}\right)^{4}\log\frac{\varepsilon_F}{T} \label{odd}\\
%c_{2n}&{=}0.20|u|^{2}\rho^{2}(\varepsilon_F)\frac{\varepsilon_F}{\hbar}\left(\frac{T}{\varepsilon_F}\right)^{2}\label{even}
%\end{align}
%for $n\geq1$, where $\rho(\varepsilon_F)=\frac{1}{2\pi\left(\hbar v_{F}\right) ^{2}}$ is the density of states (states per unit area and energy) of graphene per spin and valley at $\varepsilon_F$. 

In the presence of interactions, the  conductance per unit width is reduced to
\begin{equation}
G=G_{\rm ball}\left[1- 1.19\frac{T}{\varepsilon_F}\sqrt{\log\left(\frac{\varepsilon_F}{T}\right)}\right],   \label{G4}
\end{equation}
where the second term describes the finite-temperature correction due to inelastic {\em e-e} scattering. 
The logarithmic tail at low temperatures is an inherent hallmark of 2D electron transport and originates from the concurrent effects of a planar geometry and conservation laws \cite{b4,b12,b13,b14,b50}. 
We point out that this correction has stronger temperature dependence than the analytic correction $\sim T^2$ found in 1D \cite{b1}, because there are fewer restrictions on the allowed scattering processes in 2D. \par

\subsection{Experimental perspective}
These results call for an experimental verification as there is hardly any measurements yet which unveil the effect of interactions on the conductance of 2D systems. %\MM{an interesting exception being the recent work by Morpurgo. } 
The main hindrance to explore this phenomenology is the momentum dissipation by impurities, phonons, and umklapp scattering. Of course, these are inevitable in any real system. However, a pragmatic requirement for the applicability of our analysis to a given system is that the momentum-conserving scattering processes be faster than all other scattering mechanisms, which opens a window to observe purely collision-dominated transport. 

The pioneering work in this direction was done by Jong and Molenkamp \cite{b55,prbjong} on (Al,Ga)As in 1995 where they used a dc current to induce a desired increase in the {\em e-e} scattering rate at $T\sim 2{\rm K}$ and investigated the Gurzhi effect. With the advent of graphene and the possibility of making very clean samples, a new arena has opened for studying collision-dominated transport. References~[\onlinecite{b9,b16,b37}] have reported mobilities of the order of 200 000 ${\rm cm^2/(Vs)}$ in suspended graphene of micron size nearly a decade ago. Most recent experiments performed on freestanding pristine graphene samples \cite{b46,b56} with mobilities of the order of $10^5 {\rm cm^2/(Vs)}$ have exhibited signatures of viscous flow. In these samples, the measured mean free path at large doping reaches a few microns while the inelastic scattering length decreases to ${\rm{100-300 nm}}$ at temperatures $T\geq 150{\rm K}$. References~[\onlinecite{b61}] and~[\onlinecite{b57}] reported the signatures of viscous flow at room temperature in samples of a graphene sheet sandwiched between hexagonal boron nitride (hBN) slabs where the inelastic mean free path is shorter than ${\rm400nm}$ in a wide range of densities and temperatures $T\geq 150{\rm K}$. These recent experiments in graphene all focus on anomalous effects of viscosity on transport in restricted geometries, which are beyond the scope of the present article. However, they have unambiguously demonstrated that there is a wide parameter regime in which electrons in graphene behave as a viscous hydrodynamic fluid~\cite{b44,b61,b45,b46,b56,b61,Moll1061,prbshaffique}; hence, this provides a  remarkable opportunity for the experimental verification of our results.

\section{Role of spatial dimension}\label{RD}
In the previous section, we have presented the leading-order correction to conductance due to {\em e-e} interactions at low temperature; see Eq. (\ref{G4}). We emphasize that this result is valid only in 2D, as any lower or higher dimensional system has drastically different relaxation dynamics. In 2D systems, energy relaxation occurs by logarithmically enhanced forward scattering  whereas  relaxation of the angular quasiparticle distribution proceeds mainly by head-on collisions. In this situation, the relaxation of odd angular modes is suppressed with a higher power of $\frac{T}{\varepsilon_F}$ in comparison with even angular modes (see Sec. \ref{RM}), which in turn are logarithmically suppressed as compared to energy relaxation~\cite{b19,b20,b21}. 

In contrast, in 1D Galilean invariant systems \cite{b1,b2,b3}, equilibration to leading order involves three-particle scattering, as opposed to the prevalent two-particle scattering mechanisms (head-on or forward scattering) in 2D. The three-particle scattering changes the number of right-going particles $N_{R}$ which in turn relaxes their energy $E_{R}$, in addition to momentum, according to the relation $d{E}_{R}/dt=-\mu d{N}_{R}/dt$ \cite{b2}. Another essential aspect that distinguishes the 1D current relaxation mechanism from the 2D case is the fact that the transfer of right movers to left movers requires the intermediate creation of a hole at the bottom of the band, which is backscattered when electrons near $\varepsilon_F$ shift from right to left movers \cite{b1,b24}. As the probability of such a high-energy hole is exponentially small at low temperatures, the corresponding equilibration length is exponentially large $\sim e^{\varepsilon_F/T}$, as discussed in Refs.~[\onlinecite{b2}] and~[\onlinecite{b3}]. This contrasts with the much milder power law growth of the relaxation length in 2D [see Eq.~(\ref{inel})]. Hence, we infer that as spatial dimensions are increased from 1D to 2D, the current relaxation is enhanced, which results in a stronger $T$ dependence of the correction to the conductance. 

In 1D, the crossover from ballistic to collision-dominated conductance has been worked out in Ref.~[\onlinecite{b2}]. Our formalism allows us to extract an analogous crossover in 2D, by solving the boundary value problem at finite length.
%While in 2D, less constraints on the scattering processes make the analysis of crossover far more complex, hence it is still an unaccomplished task.\par

In 3D, one can equally well ask the question about collision-induced corrections to the conductance. However, unlike in 2D, in a collision, initial and final momenta do not have to lie in the same plane, even when the quasiparticles are forced to the vicinity of the Fermi surface at low temperature. Indeed, for a fixed momentum transfer $\vec{q}$, there is a one-dimensional continuous manifold of kinetically allowed scattering processes \cite{b4}. The scattering weight $\delta \psi^2$ associated to these scattering processes will generically be of order $O(1)$,  independently  of the inversion parity of the considered modes. Therefore, the relaxation of odd and even modes will be comparatively fast, unlike in 2D, where they differ parametrically. Since we expect the correction to the conductance to scale as the square root of the ratio of the relaxation rates of odd and even modes, we should expect a correction of order $O(1)$ in 3D, and thus a deviation from the ballistic conductance in long samples, even at very low temperatures. This is indeed the natural expectation one might have. The fact that in 1D and 2D the collision-dominated conductance of long samples tends nevertheless to the ballistic value instead hinges on the peculiarities of low-dimensional scattering and transport.

We also note that in 3D, the energy relaxation will not occur at a faster rate than the (angular) momentum relaxation because only 2D systems are sensitive to enhanced forward scattering, and thus the assumption that  deviations of the quasiparticle distribution is essentially  angular in nature might not be parametrically justified beyond 2D.

\section*{Conclusions}
For a time-reversal invariant system with no spin-orbit coupling, the scattering rates (and hence the collision operator in the Boltzmann equation) enjoy an inversion symmetry in momentum space. This symmetry allows us to classify deviations of the distribution function from its equilibrium according to their parity under inversion. In 2D, the even and odd modes relax at vastly different rates provided that head-on collisions are the dominant mechanism of relaxation, which holds for simply connected and convex but otherwise arbitrary 2D Fermi surfaces. In this case, there are only two kinetically allowed channels for collisions: forward scattering, which relaxes the energy distribution of quasiparticles at a logarithmically enhanced rate, and head-on collisions, which relax the angular distribution of quasiparticles, although less efficiently than the former. Odd-parity modes have a suppressed relaxation as compared to even-parity modes, and thus they live longer. This ensures that the quasiparticle distribution is very close to being odd under parity, despite being off equilibrium.
This in turn guarantees that the weight of the current-carrying mode is essentially the same (up to a corrections vanishing as $\frac{T}{\varepsilon_F}\sqrt{\log\frac{\varepsilon_F}{T}}$ at low temperature) as in a ballistic, noninteracting setting. This result is valid  as long as momentum-conserving {\em e-e} scattering processes dominate, that is, for samples longer than the inelastic relaxation length but shorter than the length scale on which subdominant scattering processes start relaxing the momentum of the electron fluid.

\section*{Outlook}
Our work can be extended to cases where the collisions are not inversion symmetric anymore.  This situation may arise due to spin orbit coupling in time-reversal invariant systems, or by explicitly breaking the time-reversal symmetry of the system. The simplest way to break time-reversal symmetry is by applying a perpendicular magnetic field \cite{b5}. As long as the cyclotron frequency is smaller than a pertinent inelastic rate, it should  only have a perturbative effect, which might nevertheless modify the $T$ dependence of the correction to the conductance in a significant way. It would be interesting to contrast low-temperature conductance measurements of systems that do or do not obey inversion or time-reversal symmetry.  

\begin{acknowledgments}
A. Uzair acknowledges support from the ICTP-IAEA STEP programme. M. M\"uller and A. Uzair  thank the Physics Department of the University of Basel for hospitality while part of this work was accomplished. The authors are grateful to Leonid Glazman for discussions on this problem.
\end{acknowledgments}
\appendix{}

\section{Relaxation rates due to collisions}\label{Relrates}
\subsection{Scattering weights}
Here we identify the dominant scattering processes that relax distortions of the distribution function. In particular, we study the eigenmodes of the collision operator in the even and odd sectors under the inversion $I$. Thereby, we concentrate on low temperatures and thus restrict the discussion to modes that only depend on the angle of $\vec{k}$ but not on its magnitude $k$, 
\bea
\label{angular modes}
\Psi(\vec{k})= \Psi(\theta).
\eea
At low temperature, kinematic constraints and Pauli blocking allow essentially only for two channels of scattering of quasiparticles with momenta close to the Fermi surface: (i) forward scattering where $(\vec{k}, \vec{k}')\to (\vec{k}+\vec{q}, \vec{k}'-\vec{q})$ with small $|\vec{q}| = O(T/\hbar v_{F})$ and (ii) head-on collisions $(\vec{k}, \vec{k}')\to (\vec{k}+\vec{q}, \vec{k}'-\vec{q})$, where now $\vec{k}'= -\vec{k}+ \delta\vec{k}'$ with small $|\delta\vec{k}'| = O(T/\hbar v_{F})$, while $\vec{q}$ is only constrained by the requirement that $\vec{k}+\vec{q}$ lie close to the Fermi surface again, with, however, $|\vec{q}| = O(k_F)$ in general.

For these two types of processes, we analyze the  "scattering weight" $\delta\Psi^2_{o,e}$ [as defined in Eq.~(\ref{scatweight})] for even and odd modes, respectively. 
Let us first consider the forward scattering processes 
\bea
\delta\Psi^2_{\rm fwd}&=&\big[\Psi(\vec{k})+\Psi(\vec {k'}) -\Psi(\vec{k}+\vec{q})-\Psi(\vec{k'}-\vec{q})\big]^2  \notag \\
%&\simeq\big[\Psi(\vec{k})+\Psi(\vec{k'}) -\Psi(\vec{k})-\vec{q}\cdot\nabla_{\vec{k}}\Psi(\vec{k}) \notag\\ 
%&-\Psi(\vec{k'})-\vec{q}\cdot\nabla_{\vec{k'}}\Psi(\vec{k'})\big]^2\notag\\
%\delta\Psi&
&\simeq & \big[\vec{q}\cdot(\nabla_{\vec{k}}\Psi(\vec{k})-\nabla_{\vec{k'}}\Psi(\vec{k'})\big]^2 \label{amp3}\\
&\sim & |\vec{q} |^2 \sim O\left( \left[\frac{T}{\hbar v_{F}}\right]^{2}\right),
\eea
whereby we used that  the natural scale of variation for modes (\ref{angular modes}) with angular dependence only is $k_F$  (not $k_{\rm th}= T/\hbar v_{F}$) and thus $T$ independent.
This forward scattering is  suppressed by the smallness of admissible momentum transfers $\vec{q}$ and leads only to rather slow angular diffusion of the quasiparticle distribution. We will see below that head-on collisions are more effective in relaxing the modes, certainly so in the case of even modes, but also for odd modes, where we will find a logarithmic enhancement as compared to the forward scattering channel.

Let us now turn to head-on collisions. The scattering weight for these processes is
\begin{align*}
\delta\Psi^{2}_{\rm ho}=&\big[\Psi(\vec{k})+\Psi( -\vec{k}+ \delta\vec{k}')-\Psi(\vec{k+q})\\
&-\Psi( -\vec{k-q}+ \delta\vec{k}')\big]^2.
%\delta\Psi&\simeq\big[\Psi(\vec{k})+\Psi(\vec{k'})+\vec{\delta\zeta_{k'}}\cdot\nabla_{\vec{k'}}\Psi(\vec{k'})\\
%&-\Psi(\vec{k+q})-\Psi(\vec{k'-q})\\ &-\vec{\delta\zeta_{k'-q}}\cdot\nabla_{\vec{k'-q}}\Psi(\vec{k'-q}) +O\left(\delta\zeta\right)^{2}\big]^2 \\
%\delta\Psi&\simeq\big[\Psi(\vec{k})+\Psi(\vec{-k})+\vec{\delta\zeta_{-k}}\cdot\nabla_{\vec{-k}}\Psi(-\vec{k})\\
%&-\Psi(\vec{k+q})-\Psi(\vec{-k-q})\\
%&-\vec{\delta\zeta_{-k-q}}\cdot\nabla_{\vec{-k-q}}\Psi(\vec{-k-q})+O\left(\delta\zeta\right)^{2}\big]^2
\end{align*}
For even modes, we have $\Psi_e(\vec{k})=\Psi_e(-\vec{k})$, and the scattering weight evaluates is finite in the limit $T\to 0$, 
\bea
\delta\Psi^2_{e, {\rm ho}} \simeq\big[2\Psi(\vec{k})-2\Psi(\vec{k+q}) \big]^2=O(1).  \label{amp1} 
%-\vec{\delta\zeta_{-k}}\cdot\nabla_{\vec{k}}\Psi(\vec{k}) \notag \\
%&+\vec{\delta\zeta_{-k-q}}\cdot\nabla_{\vec{k+q}}\Psi(\vec{k+q})
\eea
For odd modes, however, we have $\Psi_o(\vec{k})=-\Psi_o(-\vec{k})$, and the scattering weight is suppressed
\bea
\delta\Psi^2_{o, {\rm ho}}&{\simeq} &\big[\delta\vec{k}' \cdot[\nabla_{\vec{k}}\Psi(\vec{k})-\nabla_{\vec{k+q}}\Psi(\vec{k+q})]\big]^2 \nonumber\\
&\sim &| \delta\vec{k}'|^2 =O\left( \left[\frac{T}{\hbar v_{F}}\right]^{2}\right). \label{amp2}
\eea
Odd modes thus relax substantially more slowly than even modes \cite{odd}. 

\subsection{Temperature dependence of relaxation rates}
\label{TDRR}
We are primarily interested in the temperature dependence of the relaxation rates, and their scaling as powers of $T$. The relaxation rate of an eigenmode $\Phi_n(\vec{k})$ of the collision operator is defined as the corresponding eigenvalue and can be written as 
\bea
\label{rateApp}
c_m =\frac{\left\langle\Phi_{m}(\vec{k})|C\Phi_{m}(\vec{k})\right\rangle }{\left\langle\Phi_{m}(\vec{k})|\Phi_{m}(\vec{k})\right\rangle },
\eea
where the matrix elements of $C$ can be expressed as in Eq. (\ref{rate}), containing the scattering weight $\delta\psi^2$. 
We recall that we now restrict the discussion to slowly relaxing angular modes, approximating
$\Phi_m({\vec k}) = \Phi_m(\theta)$ within the thermal window, and exponentially falling off for $||k|-k_F|> T/\hbar v_F$, similarly as in  (\ref{Phin}). We choose the normalization condition $\frac{1}{\pi}\int d\theta |\Phi_m(\theta)|^2=1$.
 Inspecting (\ref{rotinvBEq}) and (\ref{ctilde}), we see that what really matters for our problem is not the relaxation rate (\ref{rateApp}) but rather the inverse relaxation length scale
\bea
\label{realrate}
c'_m =\frac{2\left\langle\Phi_{m}(\vec{k})|C\Phi_{m}(\vec{k})\right\rangle }{v_F B\pi},
\eea
where $B$ was defined in (\ref{B}). However, since for normalized modes $\left\langle\Phi_{m}(\vec{k})|\Phi_{m}(\vec{k})\right\rangle \sim T \sim B$, $c'_m$ and $c_m$ scale the same way with $T$.

The scaling with $T$ has two main sources, standard phase-space restrictions, and the scaling of the scattering weight with $T$.
Kinematic restrictions~\cite{b4} %\cite{b18} 
for head-on collisions leave a phase-space volume that scales as $({T}/{\varepsilon_F})^2$, reflecting the volume available to choose the two-dimensional vector $\delta\vec{k}'$ of modulus or order $O(T/\hbar v_{F})$, while the remaining degree of freedom, the scattering angle, is not restricted by temperature. This phase-space volume is then multiplied by the scattering weight $\delta\Psi^2$ to yield the scaling of the relaxation rate of the considered mode. For even modes, we thus find a scattering rate
\bea
c_e \sim\frac{\varepsilon_F}{\hbar}\left( \frac{T}{\varepsilon_F}\right)^{2}. \label{ce}
\eea
Odd modes are instead suppressed by the scattering weight, as we saw above. Naively, this suggests  a scaling $c_o\sim T^4$ \cite{odd}. However, a more careful analysis shows that 
\bea
c_o \sim\frac{\varepsilon_F}{\hbar}\left(\frac{T}{\varepsilon_F}\right)^{4}\log\frac{\varepsilon_F}{T}.   \label{co}
\eea
The extra logarithmic factor is due to a logarithmic divergence in the integral over the scattering angle $\phi$ enclosed by $\vec{k}$ and $\vec{p=k+q}$.
Indeed, when the scattering angle becomes small, the phase-space volume for choosing the tangential component of $\delta \vec{k}'$ scales as $T/\phi$ as long as $\phi$ is sufficiently bigger than $T/\varepsilon_F$. The latter provides a regularizing cut-off. This effect causes a logarithmic enhancement of the odd mode relaxation rates, $c_o$, upon integration over $\phi$. In contrast, for even modes, small angle scattering is not  beneficial  because the decrease of the scattering weight overcompensates the increase in phase space. Therefore, the logarithmic enhancement only appears in the relaxation rate of the odd modes.

\subsection{Logarithmic enhancement of the relaxation rate of odd modes, $c_{o}$} \label{log}
Let us analyze the relaxation rate of the modes as defined in Eq.~(\ref{rateApp}). For simplicity, we assume a rotationally invariant Fermi surface and illustrate the effect of enhanced scattering for modes $\Phi_n(\vec{k})=\cos(n\theta)$ with integer $n$; however, the logarithmic enhancement holds much more generally. 

Let us write the momentum $\vec{k}$ as $\vec{k}=k_F (1+\delta k)(\cos\theta_k,\sin\theta_k)$, where $\delta k\sim O(\frac{T}{\varepsilon_F})$ is a dimensionless number, and analogously for $\vec{k',p,p'}$.  In head-on collisions, we have angular configurations where $\theta_{k'}{=}\theta_{k}+\pi-\Delta$, $\theta_{p}{=}\theta_{k}+\phi$, and $\theta_{p'}{=}\theta_{p}+\pi-\delta$, where $\theta_{k}$ is arbitrary and the scattering angle $\phi$ is of order $O(1)$, while the angular deviations from anticollinearity of incoming and outgoing particles are small, $\Delta,\delta{\sim}O(\frac{T}{\varepsilon_F})$. 

The central element for evaluating a matrix element of the form (\ref{rate}) in the numerator of (\ref{rateApp}) is
the  integration of the corresponding scattering weight $\delta \Phi^2$ over the angles,
\bea
\label{angleint}
\int d\theta_k \int d\delta \int d\Delta \int d\phi \delta(\vec{k+k'-p-p'}) {\delta\Phi_{\vec{k,k',p}}}^2.\quad\quad
\eea

Defining $a= \delta k- \delta k'$ and $b= \delta p- \delta p'$, for $T\ll \varepsilon_F$, the total in- and outflowing momenta can be expressed as $\vec{k+k'}=k_F(a, \Delta)^{\rm T}$ and 
\bea
\vec{p}+\vec{p}'=
k_F \left(
\begin{array}{cc}
\cos(\phi) & -\sin(\phi)\\
\sin(\phi) & \cos(\phi)
\end{array}
\right)
\left(
\begin{array}{c}
b \\ \delta
\end{array}
\right).
\eea
The momentum-conserving $\delta$ functions can then be expressed as %\delta(\hbar v_F(\delta k + \delta k'- \delta p -\delta p'))\delta()
\bea 
\label{momcons}
&&\delta(\vec{k+k'-p-p'})=\\
&& \quad\frac{1}{k_F^{2} |\sin(\phi)|} \delta\left(\delta-\frac{b\cos\phi-a}{\sin\phi}\right)\delta\left(\Delta-\frac{b-a\cos\phi}{\sin\phi}\right),\nonumber
\eea
which does not depend on $\theta_\vec{k}$. Integrating $\delta\Phi_{\vec{k,k',p}}^2= (\cos(n\theta_{k})+\cos(n\theta_{k'})-\cos(n\theta_{p})-\cos(n\theta_{p'}))^2$ over $\theta_{k}$, we find
\bea
\label{dpsi2App}
\int d \theta_k  \delta\Phi_{\vec{k,k',p}}^2 &=& 4\pi[1+(-1)^n](1-\cos n\phi) \\
&&+2\pi n[1+(-1)^n] (\Delta-\delta)\sin n\phi \notag\\
 && +\pi n^2[\delta^2-2\delta\Delta\cos n\phi +\Delta^2].\notag
\eea
Note that for even modes (even $n$) the first line dominates, while for odd ones only the last one survives, which contains two extra small factors of $\delta, \Delta=O(T/\varepsilon_F)$.

Integrating over $\Delta$ and $\delta$ in (\ref{angleint}) and using (\ref{momcons}) and (\ref{dpsi2App}), we are left with the integral
\bea
&&\int \frac{d\phi}{k_F^{2} |\sin(\phi)|} \left\{  4\pi[1+(-1)^n](1-\cos n\phi) \right.\\
 && \quad \quad +2\pi n[1+(-1)^n] (\Delta-\delta)\sin n\phi \notag\\
 && \quad \quad\left.+\pi n^2[\delta^2-2\delta\Delta\cos n\phi +\Delta^2]\notag\right\},
\eea
where $\delta$ and $\Delta$ have to be substituted with the functions of $(\phi, a,b)$ imposed by (\ref{momcons}). 
With this substitution, all terms in the parentheses behave regularly in the limit $\phi\to 0$.  However, the Jacobian factor $1/|\sin(\phi)|$ may cause a logarithmic divergence.
The dominant term for even modes is insensitive to the diverging Jacobian, which is tamed by the term $(1-\cos n\phi)$. This leads to relaxation rates of order, 
\bea
c_{n, {\rm even}} \sim T^2,
\eea 
independently of $n$, the factors of $T$ being due to the integrals over $a,b$ which is restricted by Fermi functions to $a,b,\sim T$. However, for odd modes, only the last line of the integral survives, with a finite limit of the factor $\delta^2-2\delta\Delta\cos n\phi +\Delta^2 \to n^2 (b-a)^2$ as $\phi\to 0$. With the above approximations, the integral thus diverges logarithmically. However, the divergence is actually cut off  at small angles $\phi\sim T/\varepsilon_F$, where the angular fluctuations $\Delta,\delta \sim (a,b)/\sin(\phi)$ become of order $O(1)$ and our approximation of small angles breaks down.
We thus find that in the low-temperature limit the odd modes relax with rates scaling as 
\bea
c_{n, {\rm odd}} \sim n^4 T^4\log(\varepsilon_F/T).
\eea

\bibliography{Reference}

\begin{thebibliography}{68}%
\makeatletter
\providecommand \@ifxundefined [1]{%
 \@ifx{#1\undefined}
}%
\providecommand \@ifnum [1]{%
 \ifnum #1\expandafter \@firstoftwo
 \else \expandafter \@secondoftwo
 \fi
}%
\providecommand \@ifx [1]{%
 \ifx #1\expandafter \@firstoftwo
 \else \expandafter \@secondoftwo
 \fi
}%
\providecommand \natexlab [1]{#1}%
\providecommand \enquote  [1]{``#1''}%
\providecommand \bibnamefont  [1]{#1}%
\providecommand \bibfnamefont [1]{#1}%
\providecommand \citenamefont [1]{#1}%
\providecommand \href@noop [0]{\@secondoftwo}%
\providecommand \href [0]{\begingroup \@sanitize@url \@href}%
\providecommand \@href[1]{\@@startlink{#1}\@@href}%
\providecommand \@@href[1]{\endgroup#1\@@endlink}%
\providecommand \@sanitize@url [0]{\catcode `\\12\catcode `\$12\catcode
  `\&12\catcode `\#12\catcode `\^12\catcode `\_12\catcode `\%12\relax}%
\providecommand \@@startlink[1]{}%
\providecommand \@@endlink[0]{}%
\providecommand \url  [0]{\begingroup\@sanitize@url \@url }%
\providecommand \@url [1]{\endgroup\@href {#1}{\urlprefix }}%
\providecommand \urlprefix  [0]{URL }%
\providecommand \Eprint [0]{\href }%
\providecommand \doibase [0]{http://dx.doi.org/}%
\providecommand \selectlanguage [0]{\@gobble}%
\providecommand \bibinfo  [0]{\@secondoftwo}%
\providecommand \bibfield  [0]{\@secondoftwo}%
\providecommand \translation [1]{[#1]}%
\providecommand \BibitemOpen [0]{}%
\providecommand \bibitemStop [0]{}%
\providecommand \bibitemNoStop [0]{.\EOS\space}%
\providecommand \EOS [0]{\spacefactor3000\relax}%
\providecommand \BibitemShut  [1]{\csname bibitem#1\endcsname}%
\let\auto@bib@innerbib\@empty
%</preamble>
\bibitem [{\citenamefont {Bolotin}\ \emph
  {et~al.}(2008{\natexlab{a}})\citenamefont {Bolotin}, \citenamefont {Sikes},
  \citenamefont {Hone}, \citenamefont {Stormer},\ and\ \citenamefont
  {Kim}}]{b9}%
  \BibitemOpen
  \bibfield  {author} {\bibinfo {author} {\bibfnamefont {K.~I.}\ \bibnamefont
  {Bolotin}}, \bibinfo {author} {\bibfnamefont {K.~J.}\ \bibnamefont {Sikes}},
  \bibinfo {author} {\bibfnamefont {J.}~\bibnamefont {Hone}}, \bibinfo {author}
  {\bibfnamefont {H.~L.}\ \bibnamefont {Stormer}}, \ and\ \bibinfo {author}
  {\bibfnamefont {P.}~\bibnamefont {Kim}},\ }\href {\doibase
  10.1103/PhysRevLett.101.096802} {\bibfield  {journal} {\bibinfo  {journal}
  {Phys. Rev. Lett.}\ }\textbf {\bibinfo {volume} {101}},\ \bibinfo {pages}
  {096802} (\bibinfo {year} {2008}{\natexlab{a}})}\BibitemShut {NoStop}%
\bibitem [{\citenamefont {Du}\ \emph {et~al.}(2008)\citenamefont {Du},
  \citenamefont {Skachko}, \citenamefont {Barker},\ and\ \citenamefont
  {Andrei}}]{b16}%
  \BibitemOpen
  \bibfield  {author} {\bibinfo {author} {\bibfnamefont {X.}~\bibnamefont
  {Du}}, \bibinfo {author} {\bibfnamefont {I.}~\bibnamefont {Skachko}},
  \bibinfo {author} {\bibfnamefont {A.}~\bibnamefont {Barker}}, \ and\ \bibinfo
  {author} {\bibfnamefont {E.~Y.}\ \bibnamefont {Andrei}},\ }\href {\doibase
  10.1038/nnano.2008.199} {\bibfield  {journal} {\bibinfo  {journal} {Nat.
  Nanotechnol.}\ }\textbf {\bibinfo {volume} {3}},\ \bibinfo {pages} {491}
  (\bibinfo {year} {2008})}\BibitemShut {NoStop}%
\bibitem [{\citenamefont {Bolotin}\ \emph
  {et~al.}(2008{\natexlab{b}})\citenamefont {Bolotin}, \citenamefont {Sikes},
  \citenamefont {Jiang}, \citenamefont {Klima}, \citenamefont {Fudenberg},
  \citenamefont {Hone}, \citenamefont {Kim},\ and\ \citenamefont
  {Stormer}}]{b37}%
  \BibitemOpen
  \bibfield  {author} {\bibinfo {author} {\bibfnamefont {K.~I.}\ \bibnamefont
  {Bolotin}}, \bibinfo {author} {\bibfnamefont {K.~J.}\ \bibnamefont {Sikes}},
  \bibinfo {author} {\bibfnamefont {Z.}~\bibnamefont {Jiang}}, \bibinfo
  {author} {\bibfnamefont {M.}~\bibnamefont {Klima}}, \bibinfo {author}
  {\bibfnamefont {G.}~\bibnamefont {Fudenberg}}, \bibinfo {author}
  {\bibfnamefont {J.}~\bibnamefont {Hone}}, \bibinfo {author} {\bibfnamefont
  {P.}~\bibnamefont {Kim}}, \ and\ \bibinfo {author} {\bibfnamefont {H.~L.}\
  \bibnamefont {Stormer}},\ }\href {\doibase
  https://doi.org/10.1016/j.ssc.2008.02.024} {\bibfield  {journal} {\bibinfo
  {journal} {Solid State Commun.}\ }\textbf {\bibinfo {volume} {146}},\
  \bibinfo {pages} {351} (\bibinfo {year} {2008}{\natexlab{b}})}\BibitemShut
  {NoStop}%
\bibitem [{\citenamefont {Hrostowski}\ \emph {et~al.}(1955)\citenamefont
  {Hrostowski}, \citenamefont {Morin}, \citenamefont {Geballe},\ and\
  \citenamefont {Wheatley}}]{b39}%
  \BibitemOpen
  \bibfield  {author} {\bibinfo {author} {\bibfnamefont {H.~J.}\ \bibnamefont
  {Hrostowski}}, \bibinfo {author} {\bibfnamefont {F.~J.}\ \bibnamefont
  {Morin}}, \bibinfo {author} {\bibfnamefont {T.~H.}\ \bibnamefont {Geballe}},
  \ and\ \bibinfo {author} {\bibfnamefont {G.~H.}\ \bibnamefont {Wheatley}},\
  }\href {\doibase 10.1103/PhysRev.100.1672} {\bibfield  {journal} {\bibinfo
  {journal} {Phys. Rev.}\ }\textbf {\bibinfo {volume} {100}},\ \bibinfo {pages}
  {1672} (\bibinfo {year} {1955})}\BibitemShut {NoStop}%
\bibitem [{\citenamefont {Umansky}\ \emph {et~al.}(1997)\citenamefont
  {Umansky}, \citenamefont {de~Picciotto},\ and\ \citenamefont
  {Heiblum}}]{b40}%
  \BibitemOpen
  \bibfield  {author} {\bibinfo {author} {\bibfnamefont {V.}~\bibnamefont
  {Umansky}}, \bibinfo {author} {\bibfnamefont {R.}~\bibnamefont
  {de~Picciotto}}, \ and\ \bibinfo {author} {\bibfnamefont {M.}~\bibnamefont
  {Heiblum}},\ }\href {\doibase 10.1063/1.119829} {\bibfield  {journal}
  {\bibinfo  {journal} {Appl. Phys. Lett.}\ }\textbf {\bibinfo {volume} {71}},\
  \bibinfo {pages} {683} (\bibinfo {year} {1997})}\BibitemShut {NoStop}%
\bibitem [{\citenamefont {Chan}\ \emph {et~al.}(2016)\citenamefont {Chan},
  \citenamefont {Keller}, \citenamefont {Tahhan}, \citenamefont {Li},
  \citenamefont {Romanczyk}, \citenamefont {Baars},\ and\ \citenamefont
  {Mishra}}]{b41}%
  \BibitemOpen
  \bibfield  {author} {\bibinfo {author} {\bibfnamefont {S.~H.}\ \bibnamefont
  {Chan}}, \bibinfo {author} {\bibfnamefont {S.}~\bibnamefont {Keller}},
  \bibinfo {author} {\bibfnamefont {M.}~\bibnamefont {Tahhan}}, \bibinfo
  {author} {\bibfnamefont {H.}~\bibnamefont {Li}}, \bibinfo {author}
  {\bibfnamefont {B.}~\bibnamefont {Romanczyk}}, \bibinfo {author}
  {\bibfnamefont {S.~P.~D.}\ \bibnamefont {Baars}}, \ and\ \bibinfo {author}
  {\bibfnamefont {U.~K.}\ \bibnamefont {Mishra}},\ }\href
  {http://stacks.iop.org/0268-1242/31/i=6/a=065008} {\bibfield  {journal}
  {\bibinfo  {journal} {Semicond. Sci. Technol.}\ }\textbf {\bibinfo {volume}
  {31}},\ \bibinfo {pages} {065008} (\bibinfo {year} {2016})}\BibitemShut
  {NoStop}%
\bibitem [{\citenamefont {Tsoi}\ \emph {et~al.}(1999)\citenamefont {Tsoi},
  \citenamefont {Bass},\ and\ \citenamefont {Wyder}}]{b47}%
  \BibitemOpen
  \bibfield  {author} {\bibinfo {author} {\bibfnamefont {V.~S.}\ \bibnamefont
  {Tsoi}}, \bibinfo {author} {\bibfnamefont {J.}~\bibnamefont {Bass}}, \ and\
  \bibinfo {author} {\bibfnamefont {P.}~\bibnamefont {Wyder}},\ }\href
  {\doibase 10.1103/RevModPhys.71.1641} {\bibfield  {journal} {\bibinfo
  {journal} {Rev. Mod. Phys.}\ }\textbf {\bibinfo {volume} {71}},\ \bibinfo
  {pages} {1641} (\bibinfo {year} {1999})}\BibitemShut {NoStop}%
\bibitem [{\citenamefont {van Houten}\ \emph {et~al.}(1989)\citenamefont {van
  Houten}, \citenamefont {Beenakker}, \citenamefont {Williamson}, \citenamefont
  {Broekaart}, \citenamefont {van Loosdrecht}, \citenamefont {van Wees},
  \citenamefont {Mooij}, \citenamefont {Foxon},\ and\ \citenamefont
  {Harris}}]{b48}%
  \BibitemOpen
  \bibfield  {author} {\bibinfo {author} {\bibfnamefont {H.}~\bibnamefont {van
  Houten}}, \bibinfo {author} {\bibfnamefont {C.~J.}\ \bibnamefont
  {Beenakker}}, \bibinfo {author} {\bibfnamefont {J.~G.}\ \bibnamefont
  {Williamson}}, \bibinfo {author} {\bibfnamefont {M.~E.~I.}\ \bibnamefont
  {Broekaart}}, \bibinfo {author} {\bibfnamefont {P.~H.~M.}\ \bibnamefont {van
  Loosdrecht}}, \bibinfo {author} {\bibfnamefont {B.}~\bibnamefont {van Wees}},
  \bibinfo {author} {\bibfnamefont {J.~E.}\ \bibnamefont {Mooij}}, \bibinfo
  {author} {\bibfnamefont {C.~T.}\ \bibnamefont {Foxon}}, \ and\ \bibinfo
  {author} {\bibfnamefont {J.~J.}\ \bibnamefont {Harris}},\ }\href {\doibase
  10.1103/PhysRevB.39.8556} {\bibfield  {journal} {\bibinfo  {journal} {Phys.
  Rev. B}\ }\textbf {\bibinfo {volume} {39}},\ \bibinfo {pages} {8556}
  (\bibinfo {year} {1989})}\BibitemShut {NoStop}%
\bibitem [{\citenamefont {Beenakker}\ and\ \citenamefont {van
  Houten}(1991)}]{b49}%
  \BibitemOpen
  \bibfield  {author} {\bibinfo {author} {\bibfnamefont {C.}~\bibnamefont
  {Beenakker}}\ and\ \bibinfo {author} {\bibfnamefont {H.}~\bibnamefont {van
  Houten}},\ }in\ \href {\doibase
  https://doi.org/10.1016/S0081-1947(08)60091-0} {\emph {\bibinfo {booktitle}
  {Semiconductor Heterostructures and Nanostructures}}},\ \bibinfo {series}
  {Solid State Physics}, Vol.~\bibinfo {volume} {44}\ (\bibinfo  {publisher}
  {Academic Press, San Diego, CA},\ \bibinfo {year} {1991})\ pp.\ \bibinfo
  {pages} {1-- 228}\BibitemShut {NoStop}%
\bibitem [{\citenamefont {Kotov}\ \emph {et~al.}(2012)\citenamefont {Kotov},
  \citenamefont {Uchoa}, \citenamefont {Pereira}, \citenamefont {Guinea},\ and\
  \citenamefont {Neto}}]{b6}%
  \BibitemOpen
  \bibfield  {author} {\bibinfo {author} {\bibfnamefont {V.~N.}\ \bibnamefont
  {Kotov}}, \bibinfo {author} {\bibfnamefont {B.}~\bibnamefont {Uchoa}},
  \bibinfo {author} {\bibfnamefont {V.~M.}\ \bibnamefont {Pereira}}, \bibinfo
  {author} {\bibfnamefont {F.}~\bibnamefont {Guinea}}, \ and\ \bibinfo {author}
  {\bibfnamefont {A.~H.~C.}\ \bibnamefont {Neto}},\ }\href {\doibase
  10.1103/RevModPhys.84.1067} {\bibfield  {journal} {\bibinfo  {journal} {Rev.
  Mod. Phys.}\ }\textbf {\bibinfo {volume} {84}},\ \bibinfo {pages} {1067}
  (\bibinfo {year} {2012})}\BibitemShut {NoStop}%
\bibitem [{\citenamefont {Sheehy}\ and\ \citenamefont {Schmalian}(2007)}]{b28}%
  \BibitemOpen
  \bibfield  {author} {\bibinfo {author} {\bibfnamefont {D.~E.}\ \bibnamefont
  {Sheehy}}\ and\ \bibinfo {author} {\bibfnamefont {J.}~\bibnamefont
  {Schmalian}},\ }\href {\doibase 10.1103/PhysRevLett.99.226803} {\bibfield
  {journal} {\bibinfo  {journal} {Phys. Rev. Lett.}\ }\textbf {\bibinfo
  {volume} {99}},\ \bibinfo {pages} {226803} (\bibinfo {year}
  {2007})}\BibitemShut {NoStop}%
\bibitem [{\citenamefont {Sarma}\ and\ \citenamefont {Hwang}(1999)}]{b30}%
  \BibitemOpen
  \bibfield  {author} {\bibinfo {author} {\bibfnamefont {S.~D.}\ \bibnamefont
  {Sarma}}\ and\ \bibinfo {author} {\bibfnamefont {E.~H.}\ \bibnamefont
  {Hwang}},\ }\href {\doibase 10.1103/PhysRevLett.83.164} {\bibfield  {journal}
  {\bibinfo  {journal} {Phys. Rev. Lett.}\ }\textbf {\bibinfo {volume} {83}},\
  \bibinfo {pages} {164} (\bibinfo {year} {1999})}\BibitemShut {NoStop}%
\bibitem [{\citenamefont {Sarma}\ \emph {et~al.}(2000)\citenamefont {Sarma},
  \citenamefont {Hwang},\ and\ \citenamefont {Zutic}}]{b31}%
  \BibitemOpen
  \bibfield  {author} {\bibinfo {author} {\bibfnamefont {S.~D.}\ \bibnamefont
  {Sarma}}, \bibinfo {author} {\bibfnamefont {E.~H.}\ \bibnamefont {Hwang}}, \
  and\ \bibinfo {author} {\bibfnamefont {I.}~\bibnamefont {Zutic}},\ }\href
  {\doibase https://doi.org/10.1006/spmi.2000.0877} {\bibfield  {journal}
  {\bibinfo  {journal} {Superlattices Microstruct.}\ }\textbf {\bibinfo
  {volume} {27}},\ \bibinfo {pages} {421 } (\bibinfo {year}
  {2000})}\BibitemShut {NoStop}%
\bibitem [{\citenamefont {Sarma}\ and\ \citenamefont {Hwang}(2004)}]{b32}%
  \BibitemOpen
  \bibfield  {author} {\bibinfo {author} {\bibfnamefont {S.~D.}\ \bibnamefont
  {Sarma}}\ and\ \bibinfo {author} {\bibfnamefont {E.~H.}\ \bibnamefont
  {Hwang}},\ }\href {\doibase 10.1103/PhysRevB.69.195305} {\bibfield  {journal}
  {\bibinfo  {journal} {Phys. Rev. B}\ }\textbf {\bibinfo {volume} {69}},\
  \bibinfo {pages} {195305} (\bibinfo {year} {2004})}\BibitemShut {NoStop}%
\bibitem [{\citenamefont {Gurzhi}\ \emph {et~al.}(1989)\citenamefont {Gurzhi},
  \citenamefont {Kalinenko},\ and\ \citenamefont {Kopeliovich}}]{b53}%
  \BibitemOpen
  \bibfield  {author} {\bibinfo {author} {\bibfnamefont {R.~N.}\ \bibnamefont
  {Gurzhi}}, \bibinfo {author} {\bibfnamefont {A.~N.}\ \bibnamefont
  {Kalinenko}}, \ and\ \bibinfo {author} {\bibfnamefont {A.~I.}\ \bibnamefont
  {Kopeliovich}},\ }\href
  {http://www.jetp.ac.ru/cgi-bin/e/index/e/69/4/p863?a=list} {\bibfield
  {journal} {\bibinfo  {journal} {Zh. Eksp. Teor. Fiz.}\ }\textbf {\bibinfo
  {volume} {96}},\ \bibinfo {pages} {1522} (\bibinfo {year} {1989})},\ \bibinfo
  {note} {[Sov. Phys. JETP {\bf 69}, 863 (1989)]}\BibitemShut {NoStop}%
\bibitem [{\citenamefont {Volkenshtein}\ \emph {et~al.}(1971)\citenamefont
  {Volkenshtein}, \citenamefont {Novoselov},\ and\ \citenamefont
  {Startsev}}]{b54}%
  \BibitemOpen
  \bibfield  {author} {\bibinfo {author} {\bibfnamefont {N.}~\bibnamefont
  {Volkenshtein}}, \bibinfo {author} {\bibfnamefont {V.}~\bibnamefont
  {Novoselov}}, \ and\ \bibinfo {author} {\bibfnamefont {V.}~\bibnamefont
  {Startsev}},\ }\href
  {http://www.jetp.ac.ru/cgi-bin/e/index/e/33/3/p584?a=list} {\bibfield
  {journal} {\bibinfo  {journal} {Zh. Eksp. Teor. Fiz.}\ }\textbf {\bibinfo
  {volume} {60}},\ \bibinfo {pages} {1078} (\bibinfo {year} {1971})},\ \bibinfo
  {note} {[Sov. Phys. JETP {\bf 33}, 584 (1971)]}\BibitemShut {NoStop}%
\bibitem [{\citenamefont {Gurzhi}\ and\ \citenamefont
  {Kopeliovich}(1972)}]{b33}%
  \BibitemOpen
  \bibfield  {author} {\bibinfo {author} {\bibfnamefont {R.}~\bibnamefont
  {Gurzhi}}\ and\ \bibinfo {author} {\bibfnamefont {A.}~\bibnamefont
  {Kopeliovich}},\ }\href
  {http://www.jetp.ac.ru/cgi-bin/e/index/e/34/6/p1345?a=list} {\bibfield
  {journal} {\bibinfo  {journal} {Zh. Eksp. Teor. Fiz.}\ }\textbf {\bibinfo
  {volume} {61}},\ \bibinfo {pages} {2514} (\bibinfo {year} {1972})},\ \bibinfo
  {note} {[Sov. Phys. JETP {\bf 34}, 1345 (1972)]}\BibitemShut {NoStop}%
\bibitem [{\citenamefont {Gurzhi}\ \emph {et~al.}(1982)\citenamefont {Gurzhi},
  \citenamefont {Kopeliovich},\ and\ \citenamefont {Rutkevich}}]{b34}%
  \BibitemOpen
  \bibfield  {author} {\bibinfo {author} {\bibfnamefont {R.}~\bibnamefont
  {Gurzhi}}, \bibinfo {author} {\bibfnamefont {A.}~\bibnamefont {Kopeliovich}},
  \ and\ \bibinfo {author} {\bibfnamefont {S.}~\bibnamefont {Rutkevich}},\
  }\href {http://www.jetp.ac.ru/cgi-bin/e/index/e/56/1/p159?a=list} {\bibfield
  {journal} {\bibinfo  {journal} {Zh. Eksp. Teor. Fiz.}\ }\textbf {\bibinfo
  {volume} {83}},\ \bibinfo {pages} {290} (\bibinfo {year} {1982})},\ \bibinfo
  {note} {[Sov. Phys. JETP {\bf 56}, 159 (1982)]}\BibitemShut {NoStop}%
\bibitem [{\citenamefont {Gurzhi}\ and\ \citenamefont
  {Kopeliovich}(1973)}]{b35}%
  \BibitemOpen
  \bibfield  {author} {\bibinfo {author} {\bibfnamefont {R.}~\bibnamefont
  {Gurzhi}}\ and\ \bibinfo {author} {\bibfnamefont {A.}~\bibnamefont
  {Kopeliovich}},\ }\href
  {http://www.jetp.ac.ru/cgi-bin/e/index/e/37/1/p195?a=list} {\bibfield
  {journal} {\bibinfo  {journal} {Zh. Eksp. Teor. Fiz.}\ }\textbf {\bibinfo
  {volume} {64}},\ \bibinfo {pages} {380} (\bibinfo {year} {1973})},\ \bibinfo
  {note} {[Sov. Phys. JETP {\bf 37}, 195 (1973)]}\BibitemShut {NoStop}%
\bibitem [{\citenamefont {Gurzhi}\ \emph {et~al.}(1986)\citenamefont {Gurzhi},
  \citenamefont {Kalienko}, \citenamefont {Kopeliovich},\ and\ \citenamefont
  {Rutkevich}}]{b51}%
  \BibitemOpen
  \bibfield  {author} {\bibinfo {author} {\bibfnamefont {R.}~\bibnamefont
  {Gurzhi}}, \bibinfo {author} {\bibfnamefont {A.}~\bibnamefont {Kalienko}},
  \bibinfo {author} {\bibfnamefont {A.}~\bibnamefont {Kopeliovich}}, \ and\
  \bibinfo {author} {\bibfnamefont {S.}~\bibnamefont {Rutkevich}},\ }\href
  {http://www.jetp.ac.ru/cgi-bin/e/index/e/64/2/p407?a=list} {\bibfield
  {journal} {\bibinfo  {journal} {Zh. Eksp. Teor. Fiz.}\ }\textbf {\bibinfo
  {volume} {91}},\ \bibinfo {pages} {686} (\bibinfo {year} {1986})},\ \bibinfo
  {note} {[Sov. Phys. JETP {\bf 69}, 863 (1989)]}\BibitemShut {NoStop}%
\bibitem [{\citenamefont {Gurzhi}(1965)}]{b52}%
  \BibitemOpen
  \bibfield  {author} {\bibinfo {author} {\bibfnamefont {R.}~\bibnamefont
  {Gurzhi}},\ }\href {http://www.jetp.ac.ru/cgi-bin/e/index/e/20/4/p953?a=list}
  {\bibfield  {journal} {\bibinfo  {journal} {Zh. Eksp. Teor. Fiz.}\ }\textbf
  {\bibinfo {volume} {47}},\ \bibinfo {pages} {1415} (\bibinfo {year}
  {1965})},\ \bibinfo {note} {[Sov. Phys. JETP {\bf 20}, 953
  (1965)]}\BibitemShut {NoStop}%
\bibitem [{\citenamefont {Pomerantschuk}\ and\ \citenamefont
  {Landau}(1937)}]{b38}%
  \BibitemOpen
  \bibfield  {author} {\bibinfo {author} {\bibfnamefont {I.~Y.}\ \bibnamefont
  {Pomerantschuk}}\ and\ \bibinfo {author} {\bibfnamefont {L.~D.}\ \bibnamefont
  {Landau}},\ }\href@noop {} {\bibfield  {journal} {\bibinfo  {journal} {Zh.
  Eksp. Teor. Fiz}\ }\textbf {\bibinfo {volume} {7}},\ \bibinfo {pages} {379}
  (\bibinfo {year} {1937})},\ \bibinfo {note} {[Phys. Z. Sowjetunion, {\bf 10},
  649 (1936)]}\BibitemShut {NoStop}%
\bibitem [{\citenamefont {MacDonald}\ \emph {et~al.}(1981)\citenamefont
  {MacDonald}, \citenamefont {Taylor},\ and\ \citenamefont {Geldart}}]{b62}%
  \BibitemOpen
  \bibfield  {author} {\bibinfo {author} {\bibfnamefont {A.~H.}\ \bibnamefont
  {MacDonald}}, \bibinfo {author} {\bibfnamefont {R.}~\bibnamefont {Taylor}}, \
  and\ \bibinfo {author} {\bibfnamefont {D.~J.~W.}\ \bibnamefont {Geldart}},\
  }\href {\doibase 10.1103/PhysRevB.23.2718} {\bibfield  {journal} {\bibinfo
  {journal} {Phys. Rev. B}\ }\textbf {\bibinfo {volume} {23}},\ \bibinfo
  {pages} {2718} (\bibinfo {year} {1981})}\BibitemShut {NoStop}%
\bibitem [{\citenamefont {Andreev}\ \emph {et~al.}(2011)\citenamefont
  {Andreev}, \citenamefont {Kivelson},\ and\ \citenamefont {Spivak}}]{b58}%
  \BibitemOpen
  \bibfield  {author} {\bibinfo {author} {\bibfnamefont {A.~V.}\ \bibnamefont
  {Andreev}}, \bibinfo {author} {\bibfnamefont {S.~A.}\ \bibnamefont
  {Kivelson}}, \ and\ \bibinfo {author} {\bibfnamefont {B.}~\bibnamefont
  {Spivak}},\ }\href {\doibase 10.1103/PhysRevLett.106.256804} {\bibfield
  {journal} {\bibinfo  {journal} {Phys. Rev. Lett.}\ }\textbf {\bibinfo
  {volume} {106}},\ \bibinfo {pages} {256804} (\bibinfo {year}
  {2011})}\BibitemShut {NoStop}%
\bibitem [{\citenamefont {Forcella}\ \emph {et~al.}(2014)\citenamefont
  {Forcella}, \citenamefont {Zaanen}, \citenamefont {Valentinis},\ and\
  \citenamefont {van~der Marel}}]{b59}%
  \BibitemOpen
  \bibfield  {author} {\bibinfo {author} {\bibfnamefont {D.}~\bibnamefont
  {Forcella}}, \bibinfo {author} {\bibfnamefont {J.}~\bibnamefont {Zaanen}},
  \bibinfo {author} {\bibfnamefont {D.}~\bibnamefont {Valentinis}}, \ and\
  \bibinfo {author} {\bibfnamefont {D.}~\bibnamefont {van~der Marel}},\ }\href
  {\doibase 10.1103/PhysRevB.90.035143} {\bibfield  {journal} {\bibinfo
  {journal} {Phys. Rev. B}\ }\textbf {\bibinfo {volume} {90}},\ \bibinfo
  {pages} {035143} (\bibinfo {year} {2014})}\BibitemShut {NoStop}%
\bibitem [{\citenamefont {Narozhny}\ \emph {et~al.}(2015)\citenamefont
  {Narozhny}, \citenamefont {Gornyi}, \citenamefont {Titov}, \citenamefont
  {Sch\"utt},\ and\ \citenamefont {Mirlin}}]{b60}%
  \BibitemOpen
  \bibfield  {author} {\bibinfo {author} {\bibfnamefont {B.~N.}\ \bibnamefont
  {Narozhny}}, \bibinfo {author} {\bibfnamefont {I.~V.}\ \bibnamefont
  {Gornyi}}, \bibinfo {author} {\bibfnamefont {M.}~\bibnamefont {Titov}},
  \bibinfo {author} {\bibfnamefont {M.}~\bibnamefont {Sch\"utt}}, \ and\
  \bibinfo {author} {\bibfnamefont {A.~D.}\ \bibnamefont {Mirlin}},\ }\href
  {\doibase 10.1103/PhysRevB.91.035414} {\bibfield  {journal} {\bibinfo
  {journal} {Phys. Rev. B}\ }\textbf {\bibinfo {volume} {91}},\ \bibinfo
  {pages} {035414} (\bibinfo {year} {2015})}\BibitemShut {NoStop}%
\bibitem [{\citenamefont {Ho}\ \emph {et~al.}(2018)\citenamefont {Ho},
  \citenamefont {Yudhistira}, \citenamefont {Chakraborty},\ and\ \citenamefont
  {Adam}}]{prbshaffique}%
  \BibitemOpen
  \bibfield  {author} {\bibinfo {author} {\bibfnamefont {D.~Y.~H.}\
  \bibnamefont {Ho}}, \bibinfo {author} {\bibfnamefont {I.}~\bibnamefont
  {Yudhistira}}, \bibinfo {author} {\bibfnamefont {N.}~\bibnamefont
  {Chakraborty}}, \ and\ \bibinfo {author} {\bibfnamefont {S.}~\bibnamefont
  {Adam}},\ }\href {\doibase 10.1103/PhysRevB.97.121404} {\bibfield  {journal}
  {\bibinfo  {journal} {Phys. Rev. B}\ }\textbf {\bibinfo {volume} {97}},\
  \bibinfo {pages} {121404} (\bibinfo {year} {2018})}\BibitemShut {NoStop}%
\bibitem [{\citenamefont {Moll}\ \emph {et~al.}(2016)\citenamefont {Moll},
  \citenamefont {Kushwaha}, \citenamefont {Nandi}, \citenamefont {Schmidt},\
  and\ \citenamefont {Mackenzie}}]{Moll1061}%
  \BibitemOpen
  \bibfield  {author} {\bibinfo {author} {\bibfnamefont {P.~J.~W.}\
  \bibnamefont {Moll}}, \bibinfo {author} {\bibfnamefont {P.}~\bibnamefont
  {Kushwaha}}, \bibinfo {author} {\bibfnamefont {N.}~\bibnamefont {Nandi}},
  \bibinfo {author} {\bibfnamefont {B.}~\bibnamefont {Schmidt}}, \ and\
  \bibinfo {author} {\bibfnamefont {A.~P.}\ \bibnamefont {Mackenzie}},\ }\href
  {\doibase 10.1126/science.aac8385} {\bibfield  {journal} {\bibinfo  {journal}
  {Science}\ }\textbf {\bibinfo {volume} {351}},\ \bibinfo {pages} {1061}
  (\bibinfo {year} {2016})}\BibitemShut {NoStop}%
\bibitem [{\citenamefont {Nam}\ \emph {et~al.}(2017)\citenamefont {Nam},
  \citenamefont {Ki}, \citenamefont {Delgado},\ and\ \citenamefont
  {Morpurgo}}]{b43}%
  \BibitemOpen
  \bibfield  {author} {\bibinfo {author} {\bibfnamefont {Y.}~\bibnamefont
  {Nam}}, \bibinfo {author} {\bibfnamefont {D.~K.}\ \bibnamefont {Ki}},
  \bibinfo {author} {\bibfnamefont {D.~S.}\ \bibnamefont {Delgado}}, \ and\
  \bibinfo {author} {\bibfnamefont {A.~F.}\ \bibnamefont {Morpurgo}},\ }\href
  {\doibase 10.1038/nphys4218} {\bibfield  {journal} {\bibinfo  {journal} {Nat.
  Phys.}\ }\textbf {\bibinfo {volume} {13}},\ \bibinfo {pages} {1207} (\bibinfo
  {year} {2017})}\BibitemShut {NoStop}%
\bibitem [{\citenamefont {Guo}\ \emph {et~al.}(2017)\citenamefont {Guo},
  \citenamefont {Ilseven}, \citenamefont {Falkovich},\ and\ \citenamefont
  {Levitov}}]{b44}%
  \BibitemOpen
  \bibfield  {author} {\bibinfo {author} {\bibfnamefont {H.}~\bibnamefont
  {Guo}}, \bibinfo {author} {\bibfnamefont {E.}~\bibnamefont {Ilseven}},
  \bibinfo {author} {\bibfnamefont {G.}~\bibnamefont {Falkovich}}, \ and\
  \bibinfo {author} {\bibfnamefont {L.~S.}\ \bibnamefont {Levitov}},\ }\href
  {\doibase 10.1073/pnas.1612181114} {\bibfield  {journal} {\bibinfo  {journal}
  {Proc. Nat. Acad. Sci. U.S.A.}\ }\textbf {\bibinfo {volume} {114}},\ \bibinfo
  {pages} {3068} (\bibinfo {year} {2017})}\BibitemShut {NoStop}%
\bibitem [{\citenamefont {Kumar}\ \emph {et~al.}(2017)\citenamefont {Kumar},
  \citenamefont {Bandurin}, \citenamefont {Pellegrino}, \citenamefont {Cao},
  \citenamefont {Principi}, \citenamefont {Guo}, \citenamefont {Auton},
  \citenamefont {Shalom}, \citenamefont {Ponomarenko}, \citenamefont
  {Falkovich}, \citenamefont {Watanabe}, \citenamefont {Taniguchi},
  \citenamefont {Grigorieva}, \citenamefont {Levitov}, \citenamefont {Polini},\
  and\ \citenamefont {Geim}}]{b61}%
  \BibitemOpen
  \bibfield  {author} {\bibinfo {author} {\bibfnamefont {R.~K.}\ \bibnamefont
  {Kumar}}, \bibinfo {author} {\bibfnamefont {D.~A.}\ \bibnamefont {Bandurin}},
  \bibinfo {author} {\bibfnamefont {F.~M.~D.}\ \bibnamefont {Pellegrino}},
  \bibinfo {author} {\bibfnamefont {Y.}~\bibnamefont {Cao}}, \bibinfo {author}
  {\bibfnamefont {A.}~\bibnamefont {Principi}}, \bibinfo {author}
  {\bibfnamefont {H.}~\bibnamefont {Guo}}, \bibinfo {author} {\bibfnamefont
  {G.~H.}\ \bibnamefont {Auton}}, \bibinfo {author} {\bibfnamefont {M.~B.}\
  \bibnamefont {Shalom}}, \bibinfo {author} {\bibfnamefont {L.~A.}\
  \bibnamefont {Ponomarenko}}, \bibinfo {author} {\bibfnamefont
  {G.}~\bibnamefont {Falkovich}}, \bibinfo {author} {\bibfnamefont
  {K.}~\bibnamefont {Watanabe}}, \bibinfo {author} {\bibfnamefont
  {T.}~\bibnamefont {Taniguchi}}, \bibinfo {author} {\bibfnamefont {I.~V.}\
  \bibnamefont {Grigorieva}}, \bibinfo {author} {\bibfnamefont {L.~S.}\
  \bibnamefont {Levitov}}, \bibinfo {author} {\bibfnamefont {M.}~\bibnamefont
  {Polini}}, \ and\ \bibinfo {author} {\bibfnamefont {A.~K.}\ \bibnamefont
  {Geim}},\ }\href {\doibase 10.1038/nphys4240} {\bibfield  {journal} {\bibinfo
   {journal} {Nat. Phys.}\ }\textbf {\bibinfo {volume} {13}},\ \bibinfo {pages}
  {1182} (\bibinfo {year} {2017})}\BibitemShut {NoStop}%
\bibitem [{\citenamefont {Gurzh}(1963)}]{min}%
  \BibitemOpen
  \bibfield  {author} {\bibinfo {author} {\bibfnamefont {R.~N.}\ \bibnamefont
  {Gurzh}},\ }\href {http://www.jetp.ac.ru/cgi-bin/e/index/e/17/2/p521?a=list}
  {\bibfield  {journal} {\bibinfo  {journal} {Zh. Eksp. Teor. Fiz.}\ }\textbf
  {\bibinfo {volume} {17}},\ \bibinfo {pages} {521} (\bibinfo {year} {1963})},\
  \bibinfo {note} {[Sov. Phys. JETP {\bf 44}, 771 (1963)]}\BibitemShut
  {NoStop}%
\bibitem [{\citenamefont {Gurzhi}\ \emph
  {et~al.}(1995{\natexlab{a}})\citenamefont {Gurzhi}, \citenamefont
  {Kalinenko},\ and\ \citenamefont {Kopeliovich}}]{prbgurzhi3}%
  \BibitemOpen
  \bibfield  {author} {\bibinfo {author} {\bibfnamefont {R.~N.}\ \bibnamefont
  {Gurzhi}}, \bibinfo {author} {\bibfnamefont {A.~N.}\ \bibnamefont
  {Kalinenko}}, \ and\ \bibinfo {author} {\bibfnamefont {A.~I.}\ \bibnamefont
  {Kopeliovich}},\ }\href {\doibase 10.1103/PhysRevLett.74.3872} {\bibfield
  {journal} {\bibinfo  {journal} {Phys. Rev. Lett.}\ }\textbf {\bibinfo
  {volume} {74}},\ \bibinfo {pages} {3872} (\bibinfo {year}
  {1995}{\natexlab{a}})}\BibitemShut {NoStop}%
\bibitem [{\citenamefont {Molenkamp}\ and\ \citenamefont
  {de~Jong}(1994)}]{prbjong}%
  \BibitemOpen
  \bibfield  {author} {\bibinfo {author} {\bibfnamefont {L.~W.}\ \bibnamefont
  {Molenkamp}}\ and\ \bibinfo {author} {\bibfnamefont {M.~J.~M.}\ \bibnamefont
  {de~Jong}},\ }\href {\doibase 10.1103/PhysRevB.49.5038} {\bibfield  {journal}
  {\bibinfo  {journal} {Phys. Rev. B}\ }\textbf {\bibinfo {volume} {49}},\
  \bibinfo {pages} {5038} (\bibinfo {year} {1994})}\BibitemShut {NoStop}%
\bibitem [{\citenamefont {Falkovich}\ and\ \citenamefont
  {Levitov}(2017)}]{b45}%
  \BibitemOpen
  \bibfield  {author} {\bibinfo {author} {\bibfnamefont {G.}~\bibnamefont
  {Falkovich}}\ and\ \bibinfo {author} {\bibfnamefont {L.}~\bibnamefont
  {Levitov}},\ }\href {\doibase 10.1103/PhysRevLett.119.066601} {\bibfield
  {journal} {\bibinfo  {journal} {Phys. Rev. Lett.}\ }\textbf {\bibinfo
  {volume} {119}},\ \bibinfo {pages} {066601} (\bibinfo {year}
  {2017})}\BibitemShut {NoStop}%
\bibitem [{\citenamefont {Levitov}\ and\ \citenamefont
  {Falkovich}(2016)}]{b46}%
  \BibitemOpen
  \bibfield  {author} {\bibinfo {author} {\bibfnamefont {L.}~\bibnamefont
  {Levitov}}\ and\ \bibinfo {author} {\bibfnamefont {G.}~\bibnamefont
  {Falkovich}},\ }\href {\doibase 10.1038/nphys3667} {\bibfield  {journal}
  {\bibinfo  {journal} {Nat. Phys.}\ }\textbf {\bibinfo {volume} {12}},\
  \bibinfo {pages} {672} (\bibinfo {year} {2016})}\BibitemShut {NoStop}%
\bibitem [{\citenamefont {Bandurin}\ \emph {et~al.}(2016)\citenamefont
  {Bandurin}, \citenamefont {Torre}, \citenamefont {Kumar}, \citenamefont
  {Shalom}, \citenamefont {Tomadin}, \citenamefont {Principi}, \citenamefont
  {Auton}, \citenamefont {Khestanova}, \citenamefont {Novoselov}, \citenamefont
  {Grigorieva}, \citenamefont {Ponomarenko}, \citenamefont {Geim},\ and\
  \citenamefont {Polini}}]{b56}%
  \BibitemOpen
  \bibfield  {author} {\bibinfo {author} {\bibfnamefont {D.~A.}\ \bibnamefont
  {Bandurin}}, \bibinfo {author} {\bibfnamefont {I.}~\bibnamefont {Torre}},
  \bibinfo {author} {\bibfnamefont {R.~K.}\ \bibnamefont {Kumar}}, \bibinfo
  {author} {\bibfnamefont {M.~B.}\ \bibnamefont {Shalom}}, \bibinfo {author}
  {\bibfnamefont {A.}~\bibnamefont {Tomadin}}, \bibinfo {author} {\bibfnamefont
  {A.}~\bibnamefont {Principi}}, \bibinfo {author} {\bibfnamefont {G.~H.}\
  \bibnamefont {Auton}}, \bibinfo {author} {\bibfnamefont {E.}~\bibnamefont
  {Khestanova}}, \bibinfo {author} {\bibfnamefont {K.}~\bibnamefont
  {Novoselov}}, \bibinfo {author} {\bibfnamefont {I.~V.}\ \bibnamefont
  {Grigorieva}}, \bibinfo {author} {\bibfnamefont {L.~A.}\ \bibnamefont
  {Ponomarenko}}, \bibinfo {author} {\bibfnamefont {A.~K.}\ \bibnamefont
  {Geim}}, \ and\ \bibinfo {author} {\bibfnamefont {M.}~\bibnamefont
  {Polini}},\ }\href {\doibase 10.1126/science.aad0201} {\bibfield  {journal}
  {\bibinfo  {journal} {Science}\ }\textbf {\bibinfo {volume} {351}},\ \bibinfo
  {pages} {1055} (\bibinfo {year} {2016})}\BibitemShut {NoStop}%
\bibitem [{\citenamefont {Torre}\ \emph {et~al.}(2015)\citenamefont {Torre},
  \citenamefont {Tomadin}, \citenamefont {Geim},\ and\ \citenamefont
  {Polini}}]{b57}%
  \BibitemOpen
  \bibfield  {author} {\bibinfo {author} {\bibfnamefont {I.}~\bibnamefont
  {Torre}}, \bibinfo {author} {\bibfnamefont {A.}~\bibnamefont {Tomadin}},
  \bibinfo {author} {\bibfnamefont {A.~K.}\ \bibnamefont {Geim}}, \ and\
  \bibinfo {author} {\bibfnamefont {M.}~\bibnamefont {Polini}},\ }\href
  {\doibase 10.1103/PhysRevB.92.165433} {\bibfield  {journal} {\bibinfo
  {journal} {Phys. Rev. B}\ }\textbf {\bibinfo {volume} {92}},\ \bibinfo
  {pages} {165433} (\bibinfo {year} {2015})}\BibitemShut {NoStop}%
\bibitem [{\citenamefont {Kashuba}(2008)}]{b8}%
  \BibitemOpen
  \bibfield  {author} {\bibinfo {author} {\bibfnamefont {A.~B.}\ \bibnamefont
  {Kashuba}},\ }\href {\doibase 10.1103/PhysRevB.78.085415} {\bibfield
  {journal} {\bibinfo  {journal} {Phys. Rev. B}\ }\textbf {\bibinfo {volume}
  {78}},\ \bibinfo {pages} {085415} (\bibinfo {year} {2008})}\BibitemShut
  {NoStop}%
\bibitem [{\citenamefont {Fritz}\ \emph {et~al.}(2008)\citenamefont {Fritz},
  \citenamefont {Schmalian}, \citenamefont {M\"uller},\ and\ \citenamefont
  {Sachdev}}]{b10}%
  \BibitemOpen
  \bibfield  {author} {\bibinfo {author} {\bibfnamefont {L.}~\bibnamefont
  {Fritz}}, \bibinfo {author} {\bibfnamefont {J.}~\bibnamefont {Schmalian}},
  \bibinfo {author} {\bibfnamefont {M.}~\bibnamefont {M\"uller}}, \ and\
  \bibinfo {author} {\bibfnamefont {S.}~\bibnamefont {Sachdev}},\ }\href
  {\doibase 10.1103/PhysRevB.78.085416} {\bibfield  {journal} {\bibinfo
  {journal} {Phys. Rev. B}\ }\textbf {\bibinfo {volume} {78}},\ \bibinfo
  {pages} {085416} (\bibinfo {year} {2008})}\BibitemShut {NoStop}%
\bibitem [{\citenamefont {M\"uller}\ \emph {et~al.}(2009)\citenamefont
  {M\"uller}, \citenamefont {Br\"auninger},\ and\ \citenamefont
  {Trauzettel}}]{b7}%
  \BibitemOpen
  \bibfield  {author} {\bibinfo {author} {\bibfnamefont {M.}~\bibnamefont
  {M\"uller}}, \bibinfo {author} {\bibfnamefont {M.}~\bibnamefont
  {Br\"auninger}}, \ and\ \bibinfo {author} {\bibfnamefont {B.}~\bibnamefont
  {Trauzettel}},\ }\href {\doibase 10.1103/PhysRevLett.103.196801} {\bibfield
  {journal} {\bibinfo  {journal} {Phys. Rev. Lett.}\ }\textbf {\bibinfo
  {volume} {103}},\ \bibinfo {pages} {196801} (\bibinfo {year}
  {2009})}\BibitemShut {NoStop}%
\bibitem [{\citenamefont {Gurzhi}\ \emph
  {et~al.}(1995{\natexlab{b}})\citenamefont {Gurzhi}, \citenamefont
  {Kalinenko},\ and\ \citenamefont {Kopeliovich}}]{b21}%
  \BibitemOpen
  \bibfield  {author} {\bibinfo {author} {\bibfnamefont {R.~N.}\ \bibnamefont
  {Gurzhi}}, \bibinfo {author} {\bibfnamefont {A.~N.}\ \bibnamefont
  {Kalinenko}}, \ and\ \bibinfo {author} {\bibfnamefont {A.~I.}\ \bibnamefont
  {Kopeliovich}},\ }\href {\doibase 10.1103/PhysRevB.52.4744} {\bibfield
  {journal} {\bibinfo  {journal} {Phys. Rev. B}\ }\textbf {\bibinfo {volume}
  {52}},\ \bibinfo {pages} {4744} (\bibinfo {year}
  {1995}{\natexlab{b}})}\BibitemShut {NoStop}%
\bibitem [{\citenamefont {Datta}(1995)}]{b22}%
  \BibitemOpen
  \bibfield  {author} {\bibinfo {author} {\bibfnamefont {S.}~\bibnamefont
  {Datta}},\ }\href {\doibase 10.1017/CBO9780511805776} {\emph {\bibinfo
  {title} {Electronic Transport in Mesoscopic Systems}}},\ Cambridge Studies in
  Semiconductor Physics and Microelectronic Engineering\ (\bibinfo  {publisher}
  {Cambridge University Press, Cambridge, UK},\ \bibinfo {year}
  {1995})\BibitemShut {NoStop}%
\bibitem [{\citenamefont {Datta}(2005)}]{b23}%
  \BibitemOpen
  \bibfield  {author} {\bibinfo {author} {\bibfnamefont {S.}~\bibnamefont
  {Datta}},\ }\href
  {http://www.cambridge.org/us/academic/subjects/engineering/electronic-optoelectronic-devices-and-nanotechnology/quantum-transport-atom-transistor?format=HB&isbn=9780521631457}
  {\emph {\bibinfo {title} {Quantum Transport: Atom to Transistor}}}\ (\bibinfo
   {publisher} {Cambridge University Press, Cambridge, UK},\ \bibinfo {year}
  {2005})\BibitemShut {NoStop}%
\bibitem [{\citenamefont {Micklitz}\ \emph {et~al.}(2010)\citenamefont
  {Micklitz}, \citenamefont {Rech},\ and\ \citenamefont {Matveev}}]{b2}%
  \BibitemOpen
  \bibfield  {author} {\bibinfo {author} {\bibfnamefont {T.}~\bibnamefont
  {Micklitz}}, \bibinfo {author} {\bibfnamefont {J.}~\bibnamefont {Rech}}, \
  and\ \bibinfo {author} {\bibfnamefont {K.~A.}\ \bibnamefont {Matveev}},\
  }\href {\doibase 10.1103/PhysRevB.81.115313} {\bibfield  {journal} {\bibinfo
  {journal} {Phys. Rev. B}\ }\textbf {\bibinfo {volume} {81}},\ \bibinfo
  {pages} {115313} (\bibinfo {year} {2010})}\BibitemShut {NoStop}%
\bibitem [{\citenamefont {Rech}\ \emph {et~al.}(2009)\citenamefont {Rech},
  \citenamefont {Micklitz},\ and\ \citenamefont {Matveev}}]{b1}%
  \BibitemOpen
  \bibfield  {author} {\bibinfo {author} {\bibfnamefont {J.}~\bibnamefont
  {Rech}}, \bibinfo {author} {\bibfnamefont {T.}~\bibnamefont {Micklitz}}, \
  and\ \bibinfo {author} {\bibfnamefont {K.~A.}\ \bibnamefont {Matveev}},\
  }\href {\doibase 10.1103/PhysRevLett.102.116402} {\bibfield  {journal}
  {\bibinfo  {journal} {Phys. Rev. Lett.}\ }\textbf {\bibinfo {volume} {102}},\
  \bibinfo {pages} {116402} (\bibinfo {year} {2009})}\BibitemShut {NoStop}%
\bibitem [{\citenamefont {Ponomarenko}(1995)}]{b25}%
  \BibitemOpen
  \bibfield  {author} {\bibinfo {author} {\bibfnamefont {V.~V.}\ \bibnamefont
  {Ponomarenko}},\ }\href {\doibase 10.1103/PhysRevB.52.R8666} {\bibfield
  {journal} {\bibinfo  {journal} {Phys. Rev. B}\ }\textbf {\bibinfo {volume}
  {52}},\ \bibinfo {pages} {R8666} (\bibinfo {year} {1995})}\BibitemShut
  {NoStop}%
\bibitem [{\citenamefont {Safi}\ and\ \citenamefont {Schulz}(1995)}]{b26}%
  \BibitemOpen
  \bibfield  {author} {\bibinfo {author} {\bibfnamefont {I.}~\bibnamefont
  {Safi}}\ and\ \bibinfo {author} {\bibfnamefont {H.~J.}\ \bibnamefont
  {Schulz}},\ }\href {\doibase 10.1103/PhysRevB.52.R17040} {\bibfield
  {journal} {\bibinfo  {journal} {Phys. Rev. B}\ }\textbf {\bibinfo {volume}
  {52}},\ \bibinfo {pages} {R17040} (\bibinfo {year} {1995})}\BibitemShut
  {NoStop}%
\bibitem [{\citenamefont {Maslov}\ and\ \citenamefont {Stone}(1995)}]{b27}%
  \BibitemOpen
  \bibfield  {author} {\bibinfo {author} {\bibfnamefont {D.~L.}\ \bibnamefont
  {Maslov}}\ and\ \bibinfo {author} {\bibfnamefont {M.}~\bibnamefont {Stone}},\
  }\href {\doibase 10.1103/PhysRevB.52.R5539} {\bibfield  {journal} {\bibinfo
  {journal} {Phys. Rev. B}\ }\textbf {\bibinfo {volume} {52}},\ \bibinfo
  {pages} {R5539} (\bibinfo {year} {1995})}\BibitemShut {NoStop}%
\bibitem [{\citenamefont {Gurzh}\ \emph {et~al.}(1996)\citenamefont {Gurzh},
  \citenamefont {Kalinenko},\ and\ \citenamefont {Kopeliovich}}]{odd}%
  \BibitemOpen
  \bibfield  {author} {\bibinfo {author} {\bibfnamefont {R.~N.}\ \bibnamefont
  {Gurzh}}, \bibinfo {author} {\bibfnamefont {A.~N.}\ \bibnamefont
  {Kalinenko}}, \ and\ \bibinfo {author} {\bibfnamefont {A.~I.}\ \bibnamefont
  {Kopeliovich}},\ }\href@noop {} {\bibfield  {journal} {\bibinfo  {journal}
  {Surface Science}\ }\textbf {\bibinfo {volume} {361/362}},\ \bibinfo {pages}
  {497} (\bibinfo {year} {1996})}\BibitemShut {NoStop}%
\bibitem [{\citenamefont {Gurzhi}\ \emph {et~al.}(2003)\citenamefont {Gurzhi},
  \citenamefont {Kopeliovich}, \citenamefont {Kalinenko}, \citenamefont
  {Yanovsky}, \citenamefont {Bogachek}, \citenamefont {Landman}, \citenamefont
  {Buhmann},\ and\ \citenamefont {Molenkamp}}]{prbgurzhi1}%
  \BibitemOpen
  \bibfield  {author} {\bibinfo {author} {\bibfnamefont {R.~N.}\ \bibnamefont
  {Gurzhi}}, \bibinfo {author} {\bibfnamefont {A.~I.}\ \bibnamefont
  {Kopeliovich}}, \bibinfo {author} {\bibfnamefont {A.~N.}\ \bibnamefont
  {Kalinenko}}, \bibinfo {author} {\bibfnamefont {A.~V.}\ \bibnamefont
  {Yanovsky}}, \bibinfo {author} {\bibfnamefont {E.~N.}\ \bibnamefont
  {Bogachek}}, \bibinfo {author} {\bibfnamefont {U.}~\bibnamefont {Landman}},
  \bibinfo {author} {\bibfnamefont {H.}~\bibnamefont {Buhmann}}, \ and\
  \bibinfo {author} {\bibfnamefont {L.~W.}\ \bibnamefont {Molenkamp}},\ }\href
  {\doibase 10.1103/PhysRevB.68.165318} {\bibfield  {journal} {\bibinfo
  {journal} {Phys. Rev. B}\ }\textbf {\bibinfo {volume} {68}},\ \bibinfo
  {pages} {165318} (\bibinfo {year} {2003})}\BibitemShut {NoStop}%
\bibitem [{\citenamefont {Ziman}(1960)}]{b11}%
  \BibitemOpen
  \bibfield  {author} {\bibinfo {author} {\bibfnamefont {J.~M.}\ \bibnamefont
  {Ziman}},\ }\href@noop {} {\emph {\bibinfo {title} {Electrons and Phonons}}}\
  (\bibinfo  {publisher} {Oxford University Press, Oxfod, UK},\ \bibinfo {year}
  {1960})\ Chap.~\bibinfo {chapter} {7}\BibitemShut {NoStop}%
\bibitem [{\citenamefont {Rakyta}\ \emph {et~al.}(2010)\citenamefont {Rakyta},
  \citenamefont {Korm\'anyos},\ and\ \citenamefont {Cserti}}]{b36}%
  \BibitemOpen
  \bibfield  {author} {\bibinfo {author} {\bibfnamefont {P.}~\bibnamefont
  {Rakyta}}, \bibinfo {author} {\bibfnamefont {A.}~\bibnamefont {Korm\'anyos}},
  \ and\ \bibinfo {author} {\bibfnamefont {J.}~\bibnamefont {Cserti}},\ }\href
  {\doibase 10.1103/PhysRevB.82.113405} {\bibfield  {journal} {\bibinfo
  {journal} {Phys. Rev. B}\ }\textbf {\bibinfo {volume} {82}},\ \bibinfo
  {pages} {113405} (\bibinfo {year} {2010})}\BibitemShut {NoStop}%
\bibitem [{\citenamefont {Smith}\ and\ \citenamefont {Jensen}(1989)}]{b15}%
  \BibitemOpen
  \bibfield  {author} {\bibinfo {author} {\bibfnamefont {H.}~\bibnamefont
  {Smith}}\ and\ \bibinfo {author} {\bibfnamefont {H.~H.}\ \bibnamefont
  {Jensen}},\ }\href@noop {} {\emph {\bibinfo {title} {Transport Phenomena}}}\
  (\bibinfo  {publisher} {Oxford University Press, oxford, UK},\ \bibinfo
  {year} {1989})\BibitemShut {NoStop}%
\bibitem [{\citenamefont {Pal}\ \emph {et~al.}(2012)\citenamefont {Pal},
  \citenamefont {Yudson},\ and\ \citenamefont {Maslov}}]{b4}%
  \BibitemOpen
  \bibfield  {author} {\bibinfo {author} {\bibfnamefont {H.~K.}\ \bibnamefont
  {Pal}}, \bibinfo {author} {\bibfnamefont {V.~I.}\ \bibnamefont {Yudson}}, \
  and\ \bibinfo {author} {\bibfnamefont {D.~L.}\ \bibnamefont {Maslov}},\
  }\href {http://dx.doi.org/10.3952/lithjphys.52207} {\bibfield  {journal}
  {\bibinfo  {journal} {Lith. J. Phys.}\ }\textbf {\bibinfo {volume} {52}},\
  \bibinfo {pages} {142} (\bibinfo {year} {2012})}\BibitemShut {NoStop}%
\bibitem [{\citenamefont {Lyakhov}\ and\ \citenamefont
  {Mishchenko}(2003)}]{b19}%
  \BibitemOpen
  \bibfield  {author} {\bibinfo {author} {\bibfnamefont {A.~O.}\ \bibnamefont
  {Lyakhov}}\ and\ \bibinfo {author} {\bibfnamefont {E.~G.}\ \bibnamefont
  {Mishchenko}},\ }\href {\doibase 10.1103/PhysRevB.67.041304} {\bibfield
  {journal} {\bibinfo  {journal} {Phys. Rev. B}\ }\textbf {\bibinfo {volume}
  {67}},\ \bibinfo {pages} {041304} (\bibinfo {year} {2003})}\BibitemShut
  {NoStop}%
\bibitem [{\citenamefont {Sch\"utt}\ \emph {et~al.}(2011)\citenamefont
  {Sch\"utt}, \citenamefont {Ostrovsky}, \citenamefont {Gornyi},\ and\
  \citenamefont {Mirlin}}]{Schuett2011}%
  \BibitemOpen
  \bibfield  {author} {\bibinfo {author} {\bibfnamefont {M.}~\bibnamefont
  {Sch\"utt}}, \bibinfo {author} {\bibfnamefont {P.~M.}\ \bibnamefont
  {Ostrovsky}}, \bibinfo {author} {\bibfnamefont {I.~V.}\ \bibnamefont
  {Gornyi}}, \ and\ \bibinfo {author} {\bibfnamefont {A.~D.}\ \bibnamefont
  {Mirlin}},\ }\href {\doibase 10.1103/PhysRevB.83.155441} {\bibfield
  {journal} {\bibinfo  {journal} {Phys. Rev. B}\ }\textbf {\bibinfo {volume}
  {83}},\ \bibinfo {pages} {155441} (\bibinfo {year} {2011})}\BibitemShut
  {NoStop}%
\bibitem [{\citenamefont {Maslov}\ \emph {et~al.}(2011)\citenamefont {Maslov},
  \citenamefont {Yudson},\ and\ \citenamefont {Chubukov}}]{b18}%
  \BibitemOpen
  \bibfield  {author} {\bibinfo {author} {\bibfnamefont {D.~L.}\ \bibnamefont
  {Maslov}}, \bibinfo {author} {\bibfnamefont {V.~I.}\ \bibnamefont {Yudson}},
  \ and\ \bibinfo {author} {\bibfnamefont {A.~V.}\ \bibnamefont {Chubukov}},\
  }\href {\doibase 10.1103/PhysRevLett.106.106403} {\bibfield  {journal}
  {\bibinfo  {journal} {Phys. Rev. Lett.}\ }\textbf {\bibinfo {volume} {106}},\
  \bibinfo {pages} {106403} (\bibinfo {year} {2011})}\BibitemShut {NoStop}%
\bibitem [{\citenamefont {Laikhtman}(1992)}]{b20}%
  \BibitemOpen
  \bibfield  {author} {\bibinfo {author} {\bibfnamefont {B.}~\bibnamefont
  {Laikhtman}},\ }\href {\doibase 10.1103/PhysRevB.45.1259} {\bibfield
  {journal} {\bibinfo  {journal} {Phys. Rev. B}\ }\textbf {\bibinfo {volume}
  {45}},\ \bibinfo {pages} {1259} (\bibinfo {year} {1992})}\BibitemShut
  {NoStop}%
\bibitem [{\citenamefont {Giuliani}\ and\ \citenamefont {Quinn}(1982)}]{b12}%
  \BibitemOpen
  \bibfield  {author} {\bibinfo {author} {\bibfnamefont {G.~F.}\ \bibnamefont
  {Giuliani}}\ and\ \bibinfo {author} {\bibfnamefont {J.~J.}\ \bibnamefont
  {Quinn}},\ }\href {\doibase 10.1103/PhysRevB.26.4421} {\bibfield  {journal}
  {\bibinfo  {journal} {Phys. Rev. B}\ }\textbf {\bibinfo {volume} {26}},\
  \bibinfo {pages} {4421} (\bibinfo {year} {1982})}\BibitemShut {NoStop}%
\bibitem [{\citenamefont {Qian}\ and\ \citenamefont {Vignale}(2005)}]{b13}%
  \BibitemOpen
  \bibfield  {author} {\bibinfo {author} {\bibfnamefont {Z.}~\bibnamefont
  {Qian}}\ and\ \bibinfo {author} {\bibfnamefont {G.}~\bibnamefont {Vignale}},\
  }\href {\doibase 10.1103/PhysRevB.71.075112} {\bibfield  {journal} {\bibinfo
  {journal} {Phys. Rev. B}\ }\textbf {\bibinfo {volume} {71}},\ \bibinfo
  {pages} {075112} (\bibinfo {year} {2005})}\BibitemShut {NoStop}%
\bibitem [{\citenamefont {Hodges}\ \emph {et~al.}(1971)\citenamefont {Hodges},
  \citenamefont {Smith},\ and\ \citenamefont {Wilkins}}]{b14}%
  \BibitemOpen
  \bibfield  {author} {\bibinfo {author} {\bibfnamefont {C.}~\bibnamefont
  {Hodges}}, \bibinfo {author} {\bibfnamefont {H.}~\bibnamefont {Smith}}, \
  and\ \bibinfo {author} {\bibfnamefont {J.~W.}\ \bibnamefont {Wilkins}},\
  }\href {\doibase 10.1103/PhysRevB.4.302} {\bibfield  {journal} {\bibinfo
  {journal} {Phys. Rev. B}\ }\textbf {\bibinfo {volume} {4}},\ \bibinfo {pages}
  {302} (\bibinfo {year} {1971})}\BibitemShut {NoStop}%
\bibitem [{\citenamefont {Predel}\ \emph {et~al.}(2000)\citenamefont {Predel},
  \citenamefont {Buhmann}, \citenamefont {Molenkamp}, \citenamefont {Gurzhi},
  \citenamefont {Kalinenko}, \citenamefont {Kopeliovich},\ and\ \citenamefont
  {Yanovsky}}]{b50}%
  \BibitemOpen
  \bibfield  {author} {\bibinfo {author} {\bibfnamefont {H.}~\bibnamefont
  {Predel}}, \bibinfo {author} {\bibfnamefont {H.}~\bibnamefont {Buhmann}},
  \bibinfo {author} {\bibfnamefont {L.~W.}\ \bibnamefont {Molenkamp}}, \bibinfo
  {author} {\bibfnamefont {R.~N.}\ \bibnamefont {Gurzhi}}, \bibinfo {author}
  {\bibfnamefont {A.~N.}\ \bibnamefont {Kalinenko}}, \bibinfo {author}
  {\bibfnamefont {A.~I.}\ \bibnamefont {Kopeliovich}}, \ and\ \bibinfo {author}
  {\bibfnamefont {A.~V.}\ \bibnamefont {Yanovsky}},\ }\href {\doibase
  10.1103/PhysRevB.62.2057} {\bibfield  {journal} {\bibinfo  {journal} {Phys.
  Rev. B}\ }\textbf {\bibinfo {volume} {62}},\ \bibinfo {pages} {2057}
  (\bibinfo {year} {2000})}\BibitemShut {NoStop}%
\bibitem [{\citenamefont {Milao}\ \emph {et~al.}(2007)\citenamefont {Milao},
  \citenamefont {Wijeratne}, \citenamefont {Zhang}, \citenamefont {Coskun},
  \citenamefont {Bao},\ and\ \citenamefont {Lau}}]{b17}%
  \BibitemOpen
  \bibfield  {author} {\bibinfo {author} {\bibfnamefont {F.}~\bibnamefont
  {Milao}}, \bibinfo {author} {\bibfnamefont {S.}~\bibnamefont {Wijeratne}},
  \bibinfo {author} {\bibfnamefont {Y.}~\bibnamefont {Zhang}}, \bibinfo
  {author} {\bibfnamefont {U.~C.}\ \bibnamefont {Coskun}}, \bibinfo {author}
  {\bibfnamefont {W.}~\bibnamefont {Bao}}, \ and\ \bibinfo {author}
  {\bibfnamefont {C.~N.}\ \bibnamefont {Lau}},\ }\href {\doibase
  10.1126/science.1144359} {\bibfield  {journal} {\bibinfo  {journal}
  {Science}\ }\textbf {\bibinfo {volume} {317}},\ \bibinfo {pages} {1530}
  (\bibinfo {year} {2007})}\BibitemShut {NoStop}%
\bibitem [{\citenamefont {de~Jong}\ and\ \citenamefont
  {Molenkamp}(1995)}]{b55}%
  \BibitemOpen
  \bibfield  {author} {\bibinfo {author} {\bibfnamefont {M.~J.~M.}\
  \bibnamefont {de~Jong}}\ and\ \bibinfo {author} {\bibfnamefont {L.~W.}\
  \bibnamefont {Molenkamp}},\ }\href {\doibase 10.1103/PhysRevB.51.13389}
  {\bibfield  {journal} {\bibinfo  {journal} {Phys. Rev. B}\ }\textbf {\bibinfo
  {volume} {51}},\ \bibinfo {pages} {13389} (\bibinfo {year}
  {1995})}\BibitemShut {NoStop}%
\bibitem [{\citenamefont {Micklitz}\ \emph {et~al.}(2012)\citenamefont
  {Micklitz}, \citenamefont {Levchenko},\ and\ \citenamefont {Rosch}}]{b3}%
  \BibitemOpen
  \bibfield  {author} {\bibinfo {author} {\bibfnamefont {T.}~\bibnamefont
  {Micklitz}}, \bibinfo {author} {\bibfnamefont {A.}~\bibnamefont {Levchenko}},
  \ and\ \bibinfo {author} {\bibfnamefont {A.}~\bibnamefont {Rosch}},\ }\href
  {\doibase 10.1103/PhysRevLett.109.036405} {\bibfield  {journal} {\bibinfo
  {journal} {Phys. Rev. Let.}\ }\textbf {\bibinfo {volume} {109}},\ \bibinfo
  {pages} {036405} (\bibinfo {year} {2012})}\BibitemShut {NoStop}%
\bibitem [{\citenamefont {Lunde}\ \emph {et~al.}(2007)\citenamefont {Lunde},
  \citenamefont {Flensberg},\ and\ \citenamefont {Glazman}}]{b24}%
  \BibitemOpen
  \bibfield  {author} {\bibinfo {author} {\bibfnamefont {A.~M.}\ \bibnamefont
  {Lunde}}, \bibinfo {author} {\bibfnamefont {K.}~\bibnamefont {Flensberg}}, \
  and\ \bibinfo {author} {\bibfnamefont {L.~I.}\ \bibnamefont {Glazman}},\
  }\href {\doibase 10.1103/PhysRevB.75.245418} {\bibfield  {journal} {\bibinfo
  {journal} {Phys. Rev. B}\ }\textbf {\bibinfo {volume} {75}},\ \bibinfo
  {pages} {245418} (\bibinfo {year} {2007})}\BibitemShut {NoStop}%
\bibitem [{\citenamefont {M\"uller}\ \emph {et~al.}(2008)\citenamefont
  {M\"uller}, \citenamefont {Fritz},\ and\ \citenamefont {Sachdev}}]{b5}%
  \BibitemOpen
  \bibfield  {author} {\bibinfo {author} {\bibfnamefont {M.}~\bibnamefont
  {M\"uller}}, \bibinfo {author} {\bibfnamefont {L.}~\bibnamefont {Fritz}}, \
  and\ \bibinfo {author} {\bibfnamefont {S.}~\bibnamefont {Sachdev}},\ }\href
  {\doibase 10.1103/PhysRevB.78.115406} {\bibfield  {journal} {\bibinfo
  {journal} {Phys. Rev. B}\ }\textbf {\bibinfo {volume} {78}},\ \bibinfo
  {pages} {115406} (\bibinfo {year} {2008})}\BibitemShut {NoStop}%
\end{thebibliography}%
\end{document}